\documentclass[fleqn,usenatbib]{mnras}

\usepackage{newtxtext,newtxmath}
\usepackage[T1]{fontenc}

\DeclareRobustCommand{\VAN}[3]{#2}
\let\VANthebibliography\thebibliography
\def\thebibliography{\DeclareRobustCommand{\VAN}[3]{##3}\VANthebibliography}

\usepackage{graphicx}	% Including figure files
\usepackage{amsmath}	% Advanced maths commands
\usepackage{verbatim}
\usepackage{gensymb}

\usepackage{xcolor}
\usepackage[normalem]{ulem}
\newcommand{\rda}[1]{}
\newcommand{\aky}{}
\newcommand{\edit}{}
\newcommand\gr[1]{}

\title[Kinematics of warped discs]{Characteristics of small protoplanetary disc warps in kinematic observations}

\author[A. K. Young et al.]{
Alison K. Young,$^{1,2,3}$\thanks{E-mail:alison.young@ed.ac.uk (AKY)}
Richard Alexander$^{1}$,
Giovanni Rosotti$^{1,4}$
and Christophe Pinte$^{5}$
\\
$^{1}$School of Physics and Astronomy, University of Leicester, University Road, Leicester, LE1 7RH, UK.\\
$^{2}$SUPA, Institute for Astronomy, University of Edinburgh, Blackford Hill, Edinburgh, EH9 3HJ, UK.\\
$^{3}$Centre for Exoplanet Science, University of Edinburgh, Edinburgh, EH9 3HJ, UK.\\
$^{4}$Leiden Observatory, Leiden University, P.O. Box 9513, 2300 RA Leiden, the Netherlands.\\
$^{5}$Monash Centre for Astrophysics (MoCA) and School of Physics and Astronomy, Monash University, Clayton, Vic 3800, Australia.
}

\date{Accepted XXX. Received YYY; in original form ZZZ}

\pubyear{2021}

\begin{document}
\label{firstpage}
\pagerange{\pageref{firstpage}--\pageref{lastpage}}
\maketitle

\begin{abstract}
{\aky Many circumstellar discs appear to have misaligned central regions that give rise to shadows seen in scattered light observations}. Small warps ($<20^\circ$ misalignment) are probably more common but are also more difficult to detect than the large misalignments studied previously. We present the characteristics of CO emission that may be used to identify a small disc warp, found from synthetic $^{13}$CO maps of a model misaligned circumbinary disc.
The spectra are not symmetrical, so fitting a Keplerian model is not appropriate and can hide a warp or lead to spurious features such as spirals appearing in the residuals. We {\aky quantify the observed warp structure} by fitting sinusoids to concentric annuli of the disc. From this we can trace the radial variation of the peak velocity and of the azimuth of the peak velocity, i.e., the twist. At near face-on inclinations, these radial profiles reveal the warp structure. The twist remains detectable at moderate inclinations (${i_{\rm outer~disc}\lesssim 35^{\circ}}$) {\edit in the absence of radial flows} but the measured inclination must be accurate to $\lesssim 5^{\circ}$ to allow detection of the radial variation. {\aky The observed twist does not provide a direct measure of the warp structure because of its dependence on optical depth.} The warp causes broad asymmetries in the channel maps that span several channels and that are distinct from localised features caused by embedded planets and gravitational instability. We suspect that kinematic evidence of warps may have been missed and we suggest a few examples where the data may be revisited.

%It should be a single paragraph not more than 250 words (200 words for Letters).
%No references should appear in the abstract.
\end{abstract}


%%%%%%%%%%%%%%%%%%%%%%%%%%%%%%%%%%%%%%%%%%%%%%%%%%

%%%%%%%%%%%%%%%%% BODY OF PAPER %%%%%%%%%%%%%%%%%%

\section{Introduction}

Planets are thought to form in the dusty gas discs around protostars. Many questions remain on how exactly planet formation takes place and the evolution of protoplanetary discs is far from well understood. Protoplanetary discs are a by-product of star formation and, since most stars form in multiple systems, circumbinary discs are likely to be common. The conservation of angular momentum of the infalling gas leads to the formation of an accretion disc around the central star(s). In the simplest, and probably most common, case, the angular momentum vectors of the star(s) and disc are parallel. The disc evolves viscously and there is no out-of-plane motion, aside from internal motion of the gas within the disc due to turbulence for example and any disc winds.

Theoretical work has demonstrated that if the orbital plane of a binary is misaligned with respect to its circumbinary disc, gravitational torques can warp and tear the disc \citep{terquem1993,papaloizou1995,fragner2010,nixon2013}. Protoplanetary discs are assumed to evolve viscously, with accretion driven by the outward transport of angular momentum by viscous torques. The origin of this angular momentum transport is assumed to be turbulence, which is parameterised as a viscosity, $\alpha_{\mathrm{SS}}$ \citep{shakura1973}. 
 A perturbing force such as that of a misaligned binary may drive a warp, which propagates in the disc. The disc viscosity determines how the warp propagates. In thin, viscous discs where the disc viscosity exceeds the aspect ratio, $\alpha_{\mathrm{SS}} > h/r$, the warp propagates diffusively \citep{papaloizou1983}. For protoplanetary discs, $\alpha_{\mathrm{SS}} < h/r$ and a warp propagates via bending waves \citep{papaloizoulin1995,lubow2000}. Protoplanetary discs are most likely to fall in the latter regime since their turbulence is generally very low \citep{flaherty2015,ovelar2016,flaherty2020}.
 Current understanding of accretion mechanisms in protoplanetary discs is incomplete and we note that disc accretion may be wind driven \citep{lesur2021}. In this case, should the effective $\alpha_{\mathrm{SS}}$ be even smaller, the disc would move further into the wavelike regime in which warps are still able to form.

 {\aky The warp structure is similar whether it is caused by a perturbing star or planet, inside, outside of, or embedded within the disc. The shape of the warp is time-dependent and the exact structure of the perturbed disc depends on several factors. For a more massive perturbing companion and/or a larger misalignment between the disc and the orbit of the companion, the amplitude of the warp is greater. A greater warp amplitude can result in higher accretion rates and accretion flows onto the central star. The effective viscosity affects the communication timescale of the warp and a low viscosity may make a disc susceptible to breaking or tearing. If the precession torque of the companion on the disc exceeds the internal torques, the disc breaks into distinct planes. The effects of such large misalignments are reasonably straightforward to detect. 
 }

{\aky Many protoplanetary discs which display out-of-plane structures have now been observed. Near infra red scattered light images of protoplanetary discs show a wide variety of shadows. Broad shadows may result from a warped region and/or an inner disc(s) with a slight inclination to the outer disc (e.g. HD~143006, \citealt{benisty2018,perez2018}; HD~139614, \citealt{muro-arena2020}; GW Orionis, \citealt{kraus2020}). To cast narrow shadows across a disc requires a highly misaligned inner disc (e.g. HD~142527, \citealt{marino2015}; HD~100452, \citealt{benisty2017}).
%for example J16042165 \citep{mayama2012,pinilla2018,nealon2020b}
The difference between the orientation of an misaligned inner disc and outer disc is also clear in the observed velocity field from molecular line observations. A twisted first moment map is typically interpreted as evidence of a warp \citep{rosenfeld2012,facchini2018,zhu2019} and the origin of the observed twist or 'S'-shape in the velocity centroids is explained by \citet{casassus2015}. {\edit A twisted velocity field, however, is also produced by fast radial flows and the two scenarios may be very difficult to distinguish} \citep{rosenfeld2014}. Radial flows occur when there is a high misalignment between the  binary/planet and the disc. Several of these features have been observed in HD~142527 and were explained by the presence of a binary companion by \citet{price2018ab}. The variability of the `dipper' star AA Tau was well fit by a warped inner disc, suggesting that such variability is another indicator of a warp \citep{esau2014,alencar2010}.}

The key motivation for the quantitative detection and measurement of warped protoplanetary discs is twofold. Firstly, in the case that the perturbing bodies are known and well-characterised, measurements of the warp provide an insight into the viscosity and thermal structure of the disc, since the warp is sensitive to those properties. Measurements of warped discs would therefore be invaluable for elucidating the hydrodynamical processes driving disc evolution.
A second reason is simply that warps indicate a perturbation. For some systems, it may be possible to link a warp with a known stellar or sub-stellar companion. Otherwise, a warp may hint at a hidden planet (e.g. \citealt{nealon2018}) or past stellar fly-by (e.g. \citealt{xiang-gruess2013,cuello2019}). Various theoretical models (e.g. \citealt{terquem2013,xiang-gruess2013,nealon2018} ) have been used to examine the disc structures that would result from embedded planets of various masses and at various inclinations. While there is not a unique correlation between structure and planet, warp measurements are likely to prove useful to place constraints on the nature of possible planets and allow some scenarios to be ruled out.

% Motivation for this work 
Previous work to elucidate the observational characteristics of warped discs has tended to concentrate on those with large warps and broken discs. \citet{juhasz2017} studied simulated discs with initial misalignments of 15$^\circ$ and 30$^\circ$. Their findings include that warps cannot be detected with spatially unresolved spectra but resolved CO observations reveal brightness asymmetries and a twisted first moment map. \citet{facchini2018} show that with an initial inclination of 60$^\circ$ between the binary and disc, the resulting broken disc is readily detectable.

%Aims/Problems to address
{\aky The well--studied disc TW Hya shows signs of a small central misalignment (e.g. \citealt{roberge2005,debes2017,vanboekel2017,poteet2018}) and the possibility of the confirmation of a small warp from kinematic observations is a motivation for this work. Typically, parametric models are fitted to observational kinematic data to derive the properties of a disc. We will also examine how well the CO line emission traces the intrinsic gas velocities.}
Given how common broad shadows seem to be in protoplanetary discs, small warps are expected to be common. In this paper we use hydrodynamical and radiative transfer simulations to study discs with a central binary that is only slightly misaligned with respect to the disc. We identify the characteristics of small warps in synthetic CO emission maps and suggest a method for quantifying observations of warps. Velocity fields derived from synthetic observations via both the classic first moment method and the quadratic method of \citet{teague2018} are compared to the input model to better understand how the observed velocities relate to the intrinsic gas velocities of the source.

% Outline of paper
{\aky
In section \ref{sec:methods} we detail the methods for simulating the warped discs using hydrodynamical models and for generating the CO line maps with full radiative transfer. In \ref{sec:results}, we present the synthetic CO observations of the simulated warped discs and describe the methods for analysing the them and comparing with the hydrodynamical models. We then examine how well the CO maps trace the intrinsic disc structure and the effects of viewing inclination, uncertainty in the source inclination and the optical depth on the derived warp structure. Finally, in section \ref{sec:discussion} we discuss the kinematic characteristics of small warps and the interpretation of line observations and provide the implications for identifying small protoplanetary disc warps.}

\section{Numerical methods}
\label{sec:methods}
\subsection{Hydrodynamical simulations}

\begin{figure*}
\centering
 \includegraphics[width=0.9\textwidth,trim= 0cm 0cm 0.5cm 0cm, clip]{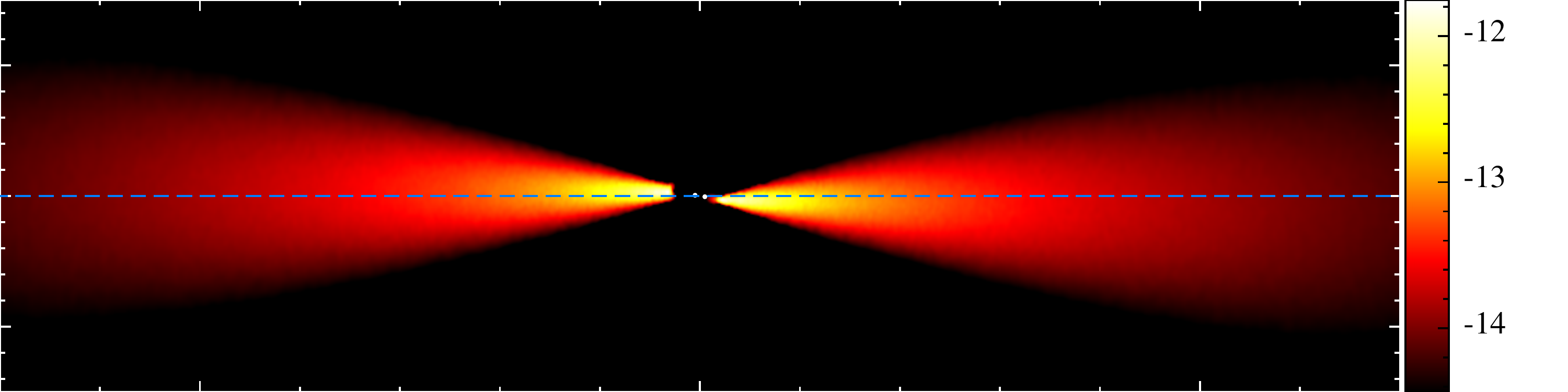}
 \caption{Density slice through the snapshot from the $10^\circ$ binary inclination hydrodynamical simulation used for the analysis here. The colour bar gives the density scale in g~cm$^{-3}$ and the dashed line indicates the initial position of the disc midplane. The vertical extend of the image is 30~au and the horizontal extent is 70~au. The vertical scale is stretched to emphasize the warp. The positions of the central stars (sink particles) are indicated by the white points.}
 \label{fig:hydroslice}
 \end{figure*}
 
Hydrodynamical modelling was performed with {\sc Phantom} \citep{price2018aa}, a smoothed particle hydrodynamics (SPH) code that been used extensively to model disc warping and tearing (e.g. \citealt{lodato2010,nixon2013,nealon2018}). While an embedded planet may arguably be the most exciting cause of warping in a protoplanetary disc, we opt to simulate a disc around a misaligned, circular equal mass binary. {\aky We would like to identify signatures of the small warp independently of any other structures. A circular binary allows us to study the disc kinematics as simply as possible, without a gap forming in the disc, a misaligned inner disc or a planet-induced spiral.}

We consider an equal mass binary modelled with sink particles \citep{bate1995aa}, which are free to accrete material from the disc. The sink particles are both 1~M$_{\odot}$, with semimajor axis $a=1$~au and have an accretion radius of 0.1~au. The disc is modelled with $5\times10^6$ SPH particles and has an initial inner radius of $r_{\mathrm{in}}=2.2$~au and outer radius $r_{\mathrm{out}}=100$~au. The aspect ratio $h/r=0.05$ at $r_{\mathrm{in}}$ and $h/r=0.13$ at $r_{\mathrm{out}}$. The surface density is initially axisymmetric and set via $\Sigma(r)=\Sigma_0 (r/r_{\mathrm{in}})^{-p}$, where $\Sigma_0=14.3$~g~cm$^{-2}$ and  $p=0.5$. Self-gravity is not included so the disc mass has no effect on the evolution of the disc or of the binary. We use a locally isothermal equation of state with $c_s \propto r^{-0.25}$.

As discussed above, the waves excited by a perturbation (in this case, the {\aky torque from the} misaligned binary) propagate via bending waves in protoplanetary discs. We therefore require to maintain $\alpha < h/r$ to maintain this regime as far as possible. The artificial numerical viscosity is set with $\beta_{\mathrm{AV}}=2$ and $\alpha_{\mathrm{AV}}$ is varied with the switch of \citet{cullen2010} between 0.01 and 1.0. Additionally, we impose $\alpha_{\textrm{SS}}=0.01$. Lower values of $\alpha_{\textrm{SS}}$ are posited for protoplanetary discs, however lower values can be  modelled reliably only with much higher resolution because of the contribution of the numerical viscosity in SPH, which is resolution-dependent. For this work, we are concerned only with the observational appearance of small warps so it is inconsequential if the warp propagation is not always wavelike in the simulations.

Three simulations are run with the binary initially inclined to the disc plane at an angle of 5, 10 and 20$^\circ$. Additionally, we run an `aligned' simulation where the binary orbits in same plane as the disc for comparison. Snapshots are extracted after 1350 binary orbits (955 years) for analysis. {\aky The disc radial communication timescale is $\sim 4600$ yr \citep{lubow2018}. We extract the snapshot before the warp has propagated to the outer edge of the disc,} preventing any erroneous reflections. The global precession time is found following the method described in {\aky \citet{smallwood2019aa} (their Eq. 6) and is 15,000 yr. The alignment timescale is significantly shorter in simulations than analytical predictions due to the additional dissipation due to the numerical viscosity. Over the course of 1600 orbits the misalignment of the binary orbit and disc in the 10$^\circ$ binary inclination simulation decreases by less than 0.8$^\circ$ at 20~au and less than 0.4$^\circ$. Alignment of the disc to the binary orbital plane is not therefore not a significant factor to consider here.  }

\subsection{Radiative transfer: synthetic line maps}
The snapshots from the hydrodynamical simulations are read into a radiative transfer code, {\sc mcfost} \citep{pinte2006,pinte2009}, to simulate CO line maps. A 3D Voronoi mesh is created from the SPH particle {\aky density} values, which is then used for the temperature and line transfer calculations. {\aky The distance to the source is set to 140~pc.} A uniform dust-to-gas ratio of 0.01 with dust grain sizes following a power law distribution $N\propto a^{-3.5}$, for grain sizes $0.03\mathrm{\mu m} < a < 1$~mm \citep{mathis1977}. For the dust optical properties we use the `smoothed astronomical silicate' model \citep{drainelee1984,laor1993,weingartner2001}. The stars each have a luminosity of 2.1 L$_\odot$, which was estimated from the 1Myr Siess isochrone \citep{siess2000}. The dust temperatures were then calculated assuming radiative equilibrium.

CO molecular line emission was then calculated {\aky using the velocity data from the hydrodynamical simulation and} assuming the dust and gas temperatures are equal. We assume local thermodynamic equilibrium. The line we choose to model, $^{13}$CO, originates deep enough in the disc for this assumption to hold. The relative abundance of $^{13}$CO is $1.3\times10^{-6}$ \citep{woods2009}. We implement an approximation for freeze out of CO, wherein the abundance is set to zero in regions where $T<20$~K. In practice, this has little effect because only cells very close to the midplane are altered. The channel width is 50~ms$^{-1}$, which is the best spectral resolution feasible for interferometric observations of a protoplanetary disc. CO maps are also simulated with the disc density scaled by factors of 0.1 and 10 to provide an approximation of the effect of the disc opacity on the observed velocity.

\begin{figure}
 \centering
  \includegraphics[width=0.8\columnwidth,trim= 1.5cm 0cm 2.5cm 1cm, clip]{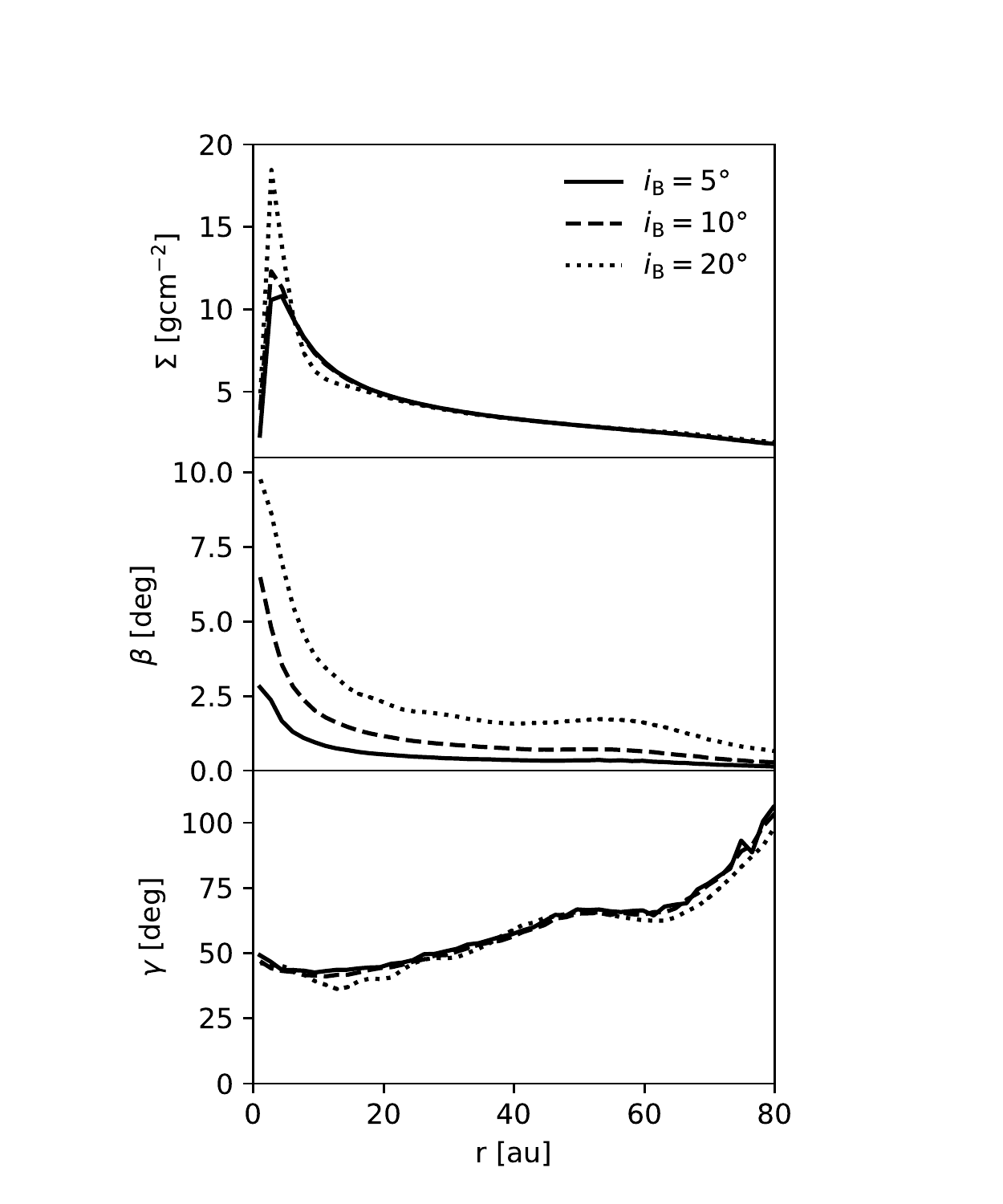}
  \caption{ Surface density ($\Sigma$), tilt ($\beta$) and twist ($\gamma$) profiles at 1350 orbits for the three simulations with the binary inclined by 5, 10 and 20\degree ~with respect to the disc. The tilt differs between the models whereas the twist is similar for all three.}
  \label{fig:angm}
\end{figure}

%%%%%% RESULTS %%%%%%%%%%%%%%%%%%%%%%%
\section{Results}
\label{sec:results}
\subsection{Hydrodynamical simulations}

The disc is initially aligned with the $x$-$y$ plane while the orbital plane of the binary is inclined with the ascending node located on the positive $x$-axis. Both the binary and disc rotation are anticlockwise. As the system evolves, a warp develops near to the inner disc boundary, deforming the centre of the disc out of the plane, and the wave propagates outwards. In the snapshots shown, the warp is oriented with the maximum extent away from the observer located $\sim 45^\circ$ anticlockwise from the $x$-axis. The disc morphology can be seen in the density slice from the $10^\circ$ binary inclination simulation presented in Fig.~\ref{fig:hydroslice}.

The angular momentum unit vector varies radially in a warped disc. The unit vector of the specific angular momentum for a ring within the disc is $\hat{l}=(\cos\gamma\sin\beta,\sin\gamma\sin\beta,\cos\beta)$. Here two angles are defined, $\beta$ and $\gamma$, which are respectively known as the tilt and twist. The tilt and twist describe the shape of the warped disc and represent the rotation relative to the $z$- and $x$- axes.

The disc surface density profile, $\Sigma (r)$, and angular momentum components tilt, $\beta$, and twist, $\gamma$, after 1350 binary orbits are plotted in Fig.~\ref{fig:angm}.
The disc surface density profile and twist are similar for all inclinations of the central binary but the tilt increases with binary inclination.
\begin{figure*}
% made with CBfigures.py
\centering
  \includegraphics[width=0.8\textwidth,trim= 3cm 10.cm 2cm 2cm, clip]{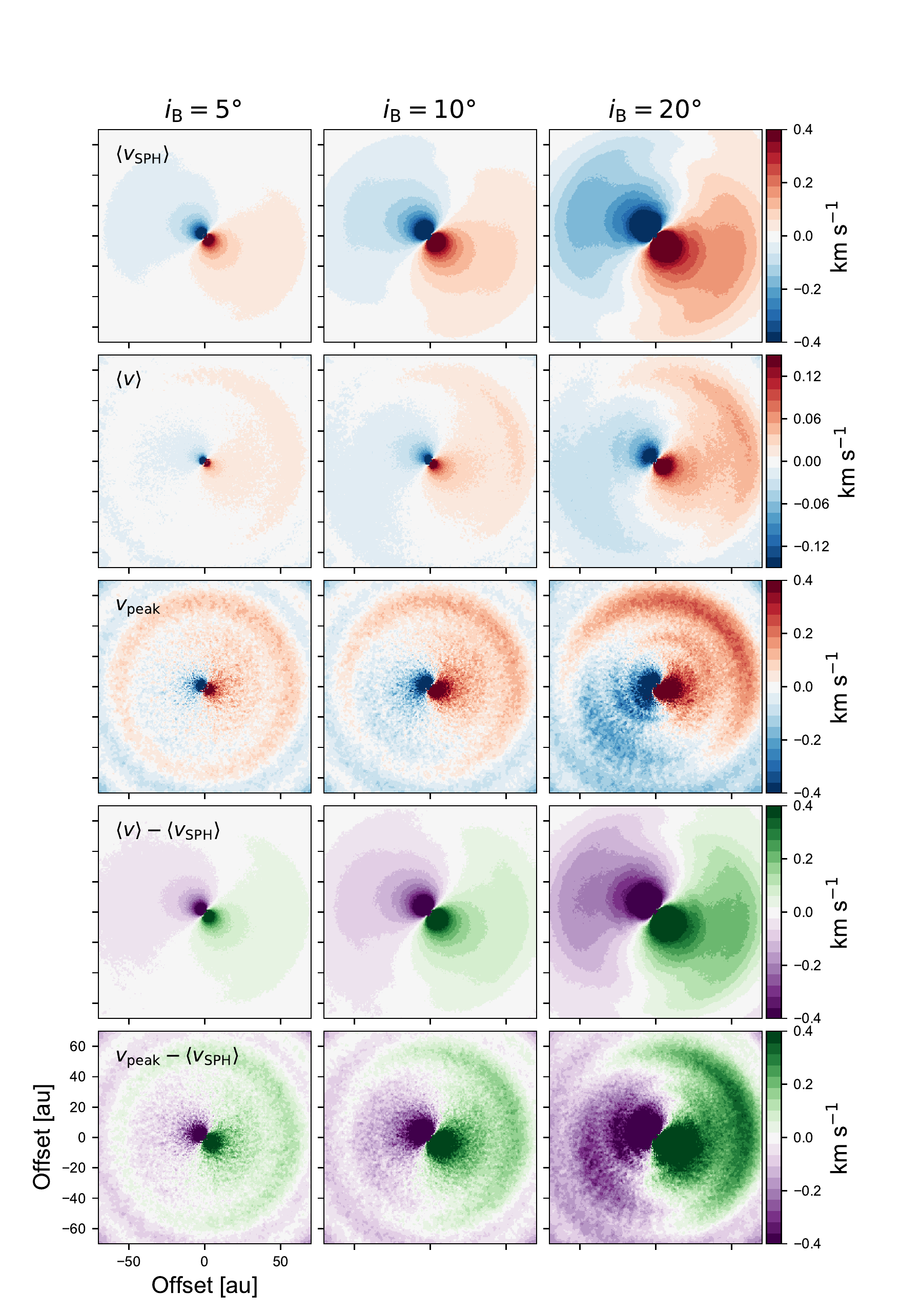}
    \caption{Velocities derived from the SPH model ($\langle v_{\rm SPH} \rangle$, density-weighted column-averaged velocity), $^{13}$CO 3-2 first moment maps,$\langle v \rangle$, and $v_{\rm{peak}}$ maps. The bottom two rows contain the residuals between the velocity fields derived from the synthetic CO maps and the raw SPH model. The columns are three simulations with the binary orbit inclined at 5, 10 and 20$^\circ$ to the disc after 1350 binary orbits. Note the different scale for the first moment maps since the observed velocities are much lower.}
   \label{fig:velocitymaps}
\end{figure*}

\subsection{Comparison of velocity fields from the hydrodynamic simulation and CO maps}
\label{sec:velcompare}

 The raw velocity field is extracted from the hydrodynamical simulation output by calculating the density-weighted average $z$-component of the velocity, $\langle v_z \rangle$, looking along the $z$ axis {\aky (the assumed line of sight here)}:
 \begin{equation}
\langle v_z \rangle = \frac{ \int{ \rho {v_z} {dz} } }{ \int{ \rho {dz} }}.
\label{eq:splashvz}
\end{equation}
 
\noindent The CO velocity maps are generated with {\sc bettermoments} \citep{teague2018,teague2019a}  from the velocity cubes. We use two methods to calculate the velocity field, the first moment map and the quadratic method {\aky of find the velocity of the peak of the spectrum}, and compare the results obtained with each. The first moment map (intensity-weighted velocity) is calculated in the standard manner:
\begin{equation}
\langle v \rangle = \frac{ \sum\limits_{i=1}^{N}{ I(v_i) {v_i} } }{ \sum\limits_{i=1}^{N}{ I(v_i) }}, 
\label{eq:moment1}
\end{equation}

\noindent where $I(v_i)$ is the flux density of velocity channel $i$. 
{\aky The peak velocity, $v_{\rm{peak}}$, map} is found by fitting a quadratic curve in velocity space to the pixels near the maximum value of the spectrum for each spatial increment in the image. The method is described in detail in \citet{teague2018}.

We concentrate on $^{13}$CO (3-2) emission, since this spectral line is commonly targeted in high-resolution submillimeter observations and will be included in the future exoALMA Large Program. Velocity maps were generated as described above for a face on ($i=0^{\circ}$) viewing inclination \footnote{\aky Here we can define the inclination as the rotation of the model from the initial orientation. In observations the inclination may be uncertain since the orientation of the mid plane varies radially in a warped disc.} of the outer disc and are shown in Fig.~\ref{fig:velocitymaps}. These velocity maps amount to a perfect deprojection and Keplerian subtraction of a disc. They show the deviation from the velocity field expected from an unperturbed Keplerian disc observed at a low inclination. Observations of inclined discs will naturally be affected by the non-uniform optical depth along the line of sight across the disc and we will discuss the implications of this later. Qualitatively, these maps reveal {\aky a change in the orientation of the red- and blue-shifted sides with radius, not dissimilar to the appearance of a misaligned inner disc.} There is a spiral tail-like feature at the outer edge of the warp wave. Similar structures are seen in the $^{13}$CO maps to the raw hydrodynamical density-weighted velocity field, $\langle v_z \rangle  $. The velocity field appears noticeably more twisted in the {\aky $v_{\rm{peak}}$ map}.

The residuals found when subtracting the SPH models' average velocity field $\langle v \rangle _{\mathrm{SPH}}$ from the synthetic CO-derived velocity fields indicate that the observed velocity fields are not equal to the `real' velocity field. There is a difference in the twist of the first moment and $v_{\rm{peak}}$ residuals and the warp appears slightly rotated in the $v_{\rm{peak}}$ map. We also see a new spiral feature appear in the residuals, which is most pronounced for the $20^\circ$ binary inclination model. We emphasise that there is no embedded planet here, but such a feature could be mistaken for a planet-induced spiral arm.

\subsection{Quantitative description of the kinematic structures}
\subsubsection{Analysis}
\label{sec:sinefit}
So far, {\aky our interpretation of kinematics observations of putative warped discs} goes little further than spotting the `twist' which makes comparisons of different discs difficult. Next, we describe the method implemented to quantify the radial dependence of the peak projected velocity and of the observed `twist'. We compare the velocity values at selected annuli centred on the image centre, which here is the centre of mass of the binary%\footnote{In the hydrodynamical model, the particles are mass-less.}
. After binning the pixels into the desired annuli, we fit a sine curve using {\sc curve\_fit} from {\sc Scipy.optimize} to the pixel azimuthal angles, $\phi$, and line-of-sight velocity values, $v_{\rm los}$, for each annulus:
\begin{equation}
v_{\rm los} = v_{\mathrm{max}} \sin \left(\phi + \theta_p\right) .
\label{eq:sinfit}
\end{equation}
{\aky An example of this fit is shown in appendix \ref{app:sinefit}}. {\aky This approach is very similar to fitting an azimuthally averaged, projected rotation curve like method presented in \citet{casassus2019a}. Here however, we fit selected radial ranges, rather than annuli covering the whole disc, and obtain two useful parameters}. $v_{\mathrm{max}}$ is the amplitude or maximum value of $v_{\rm los}$. In Fig.~\ref{fig:velocitymaps}, we saw the distinct `twist' which is already used as an observational indicator of warp structure. This twist is now quantified as $\theta_p$, the phase angle of the peak velocity. We can now compare the variation of the peak velocity and phase angle with radius for various types of observation and disc structures. 
\subsubsection{Velocity and twist profiles}
We extract velocity values for annuli at 6, 10, 12, 20, 30, 40 and 50 au, each within $\pm0.2$ au. We present the velocity radial profiles $v_{\mathrm{max}}$ and phase angle $\theta_p$ variation for the three warped disc models in Fig.~\ref{fig:CBplot}. The $v_{\mathrm{max}}$ gradients are steepest within $r \lesssim 20$ au, where the effect of the warp is greatest. The velocities remain $<1$~km~s$^{-1}$ outside 10~au, which presents a challenge for observations. For the smallest warp ($5^\circ$ binary inclination) the difference in  $v_{\mathrm{max}}$ at 50 and 10~au is $\sim 0.1$~km~s$^{-1}$ and would require very high spectral resolution to detect.

There is an offset of $\lesssim30^\circ$ between the phase angle measured from the simulation directly and from the synthetic first moment maps. For the $v_{\rm{peak}}$-derived values this increases to $\lesssim 60^\circ$. As we saw in Fig.~\ref{fig:velocitymaps}, the phase angle changes slightly faster with radius for the CO-derived values, indicating a more twisted observed structure. The change in $\theta_p$ with radius is $\sim 40^{\circ}$ between 20 and 50~au in the first moment map. The difference in $\theta_p$ from the $v_{\rm{peak}}$ maps depends on the initial binary inclination and is $\sim 30^{\circ}$ for the $10^\circ$ binary inclination model. This shift should easily be detectable in real observations. None of the observed $\theta_p$ profiles traces the sharp twist toward the centre of the disc that we see in the values measured directly from the simulation.
\subsubsection{How well does CO trace the velocity structure?}
\begin{figure*}
% made with sine_plots_CBinc.py
\centering
\includegraphics[width=\textwidth]{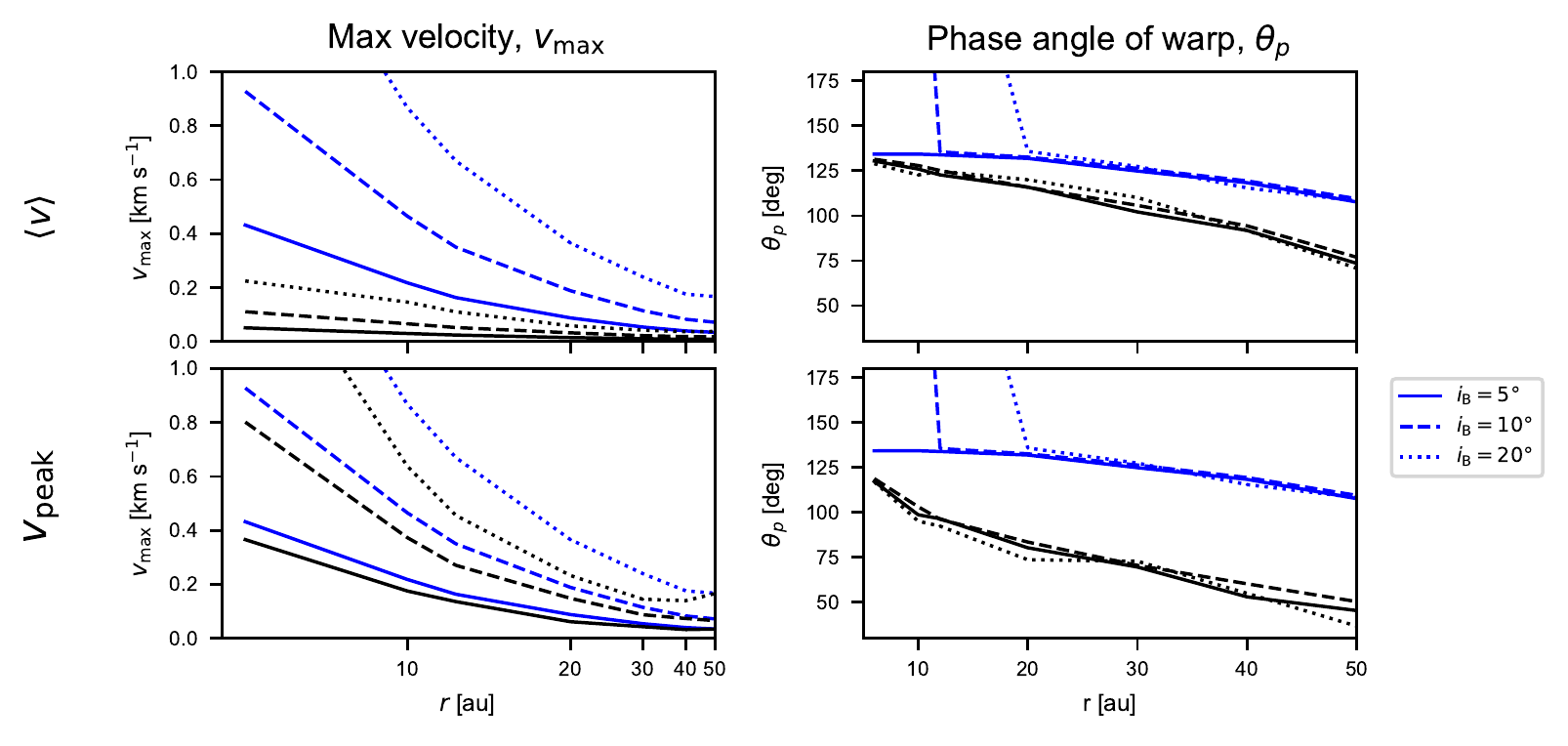}
\caption{\aky $v_{\mathrm{max}}$ and $\theta_p$ profiles for models with initial misalignments of 5, 10 and 20$^\circ$ of the binary to the disc after 1350 orbits. The profiles in the top row were derived from first moment maps and the profiles in the bottom row were derived from $v_{\rm{peak}}$ maps. Blue lines indicate results from the SPH output and black lines are results derived from the simulated $^{13}$CO maps.}
\label{fig:CBplot}
\end{figure*}

In Fig.~\ref{fig:CBplot}, the profiles extracted from synthetic CO maps are also compared to the velocity fields from the hydrodynamical simulations and this reveals how well the observationally-derived values represent the disc kinematics. 

The velocity profiles obtained from the synthetic CO maps differ markedly from the simulation velocity profile. The velocities derived from the first moment maps are smaller than those in the simulation and this difference increases at small radii and the profile is closer to a $v\propto r^{-0.5}$ Keplerian profile. The $v_{\rm{peak}}$-derived velocities are much closer to the density-weighted velocities from the simulation and follow the gradient more closely. We recall how the velocity values are calculated from the hydrodynamical file and image data cubes (section \ref{sec:velcompare}) to explain these differences. The average $z$-velocity calculated from the hydrodynamic simulation directly in Eq.~\ref{eq:splashvz} is the average value of $v_z$ in the $\hat{z}$ direction, weighted by the mass through the density, $\rho$. This is similar to the definition of the first moment map (Eq.~\ref{eq:moment1}), in which the velocity channels are weighted by the flux intensity. The residuals in Fig.~\ref{fig:velocitymaps} are greater for the $v_{\rm{peak}}$ map in places although the fitted values of $v_{\rm{max}}$ from the $v_{\rm{peak}}$ map are closer to the SPH values. This is due to the azimuthal offset of $v_{\rm{max}}$between the $v_{\rm{peak}}$ map and SPH velocity field, which increases with radius (see Fig.~\ref{fig:CBplot}, bottom right panel).

\label{app:CBincs}

\begin{figure}
\centering
\includegraphics[width=\columnwidth,trim=0cm 0cm 0cm 0cm,clip]{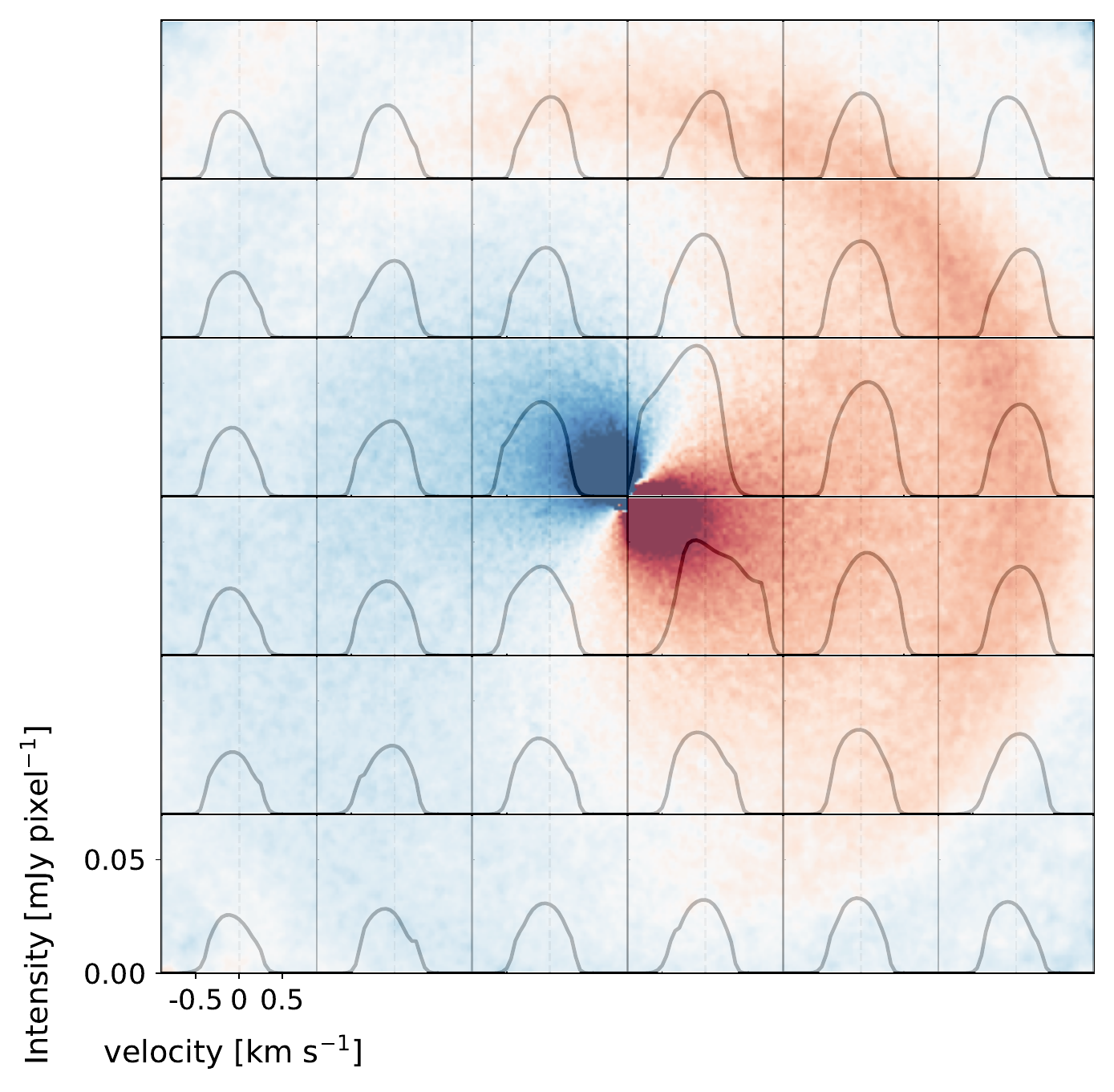}
\caption{\aky Spectra extracted from the central pixel of panels across the face-on $^{13}$CO 3-2 spectral cube for the $10^\circ$ binary inclination model after 1350 orbits and viewed face-on. Each panel is $0.17\times0.17$'' ($\sim 24\times24$~au). The spectra do not have Gaussian profiles, meaning the assumption of a Gaussian for deriving the velocity will lead to errors. In the background is the first moment map for the same cube, showing the region of the disc represented by each panel.}
\label{fig:CB10spectra}
\end{figure}

\begin{figure}
\centering
\includegraphics[width=\columnwidth,trim = 2.5cm 3.5cm 5cm 2cm, clip]{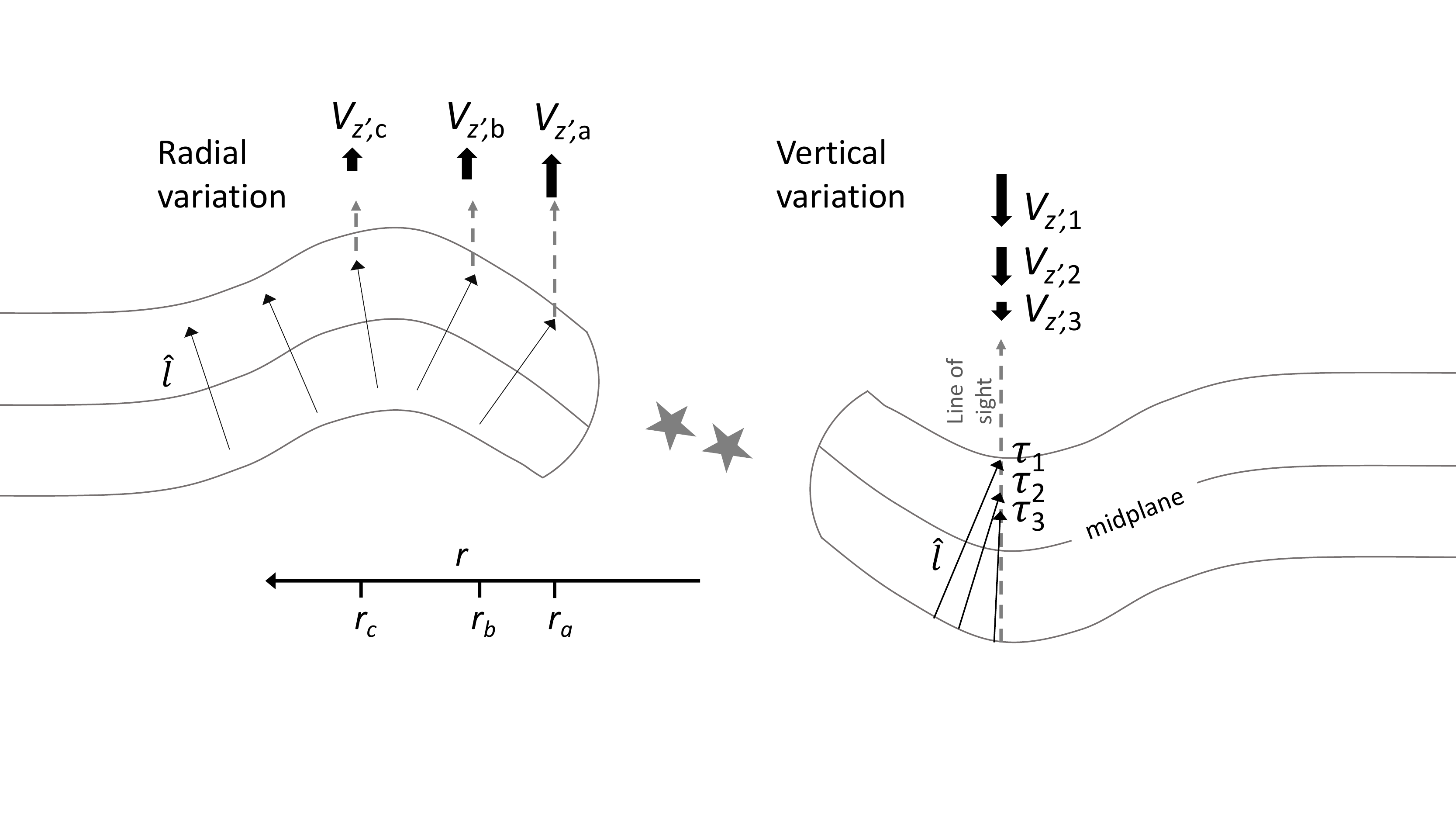}
\caption{Sketch diagram to explain the observed CO velocities. The left hand side illustrates the radial variation of the line-of-sight velocity $v_{z\prime}$ at $r_{\rm a}$, $ r_{\rm b}$ and $r_{\rm c}$. The right hand side shows the vertical variation of $v_{z\prime}$, which will be evident in tracers of differing optical depths $\tau_i$. The gas velocity is perpendicular to $\hat{l}$, the local specific angular momentum unit vector. Thin arrows indicate the orientation of $\hat{l}$ and solid arrows indicate the magnitude of the observed tangential velocity. The angular momentum vector becomes more aligned with the the line of sight deeper into the disc. Shallower layers therefore have a greater tangential velocity and greater line-of-sight velocity.}
\label{fig:diagram}
\end{figure}

The line profiles across a warped disc are not symmetrical, {\aky as would be the case for a face-on, unwarped disc}. Spectra for regions across the disc in the $10^\circ$ binary inclination model viewed face-on are shown in Fig.~\ref{fig:CB10spectra}. In many regions of the disc the peak is shifted from $v=0$ but there is also a `shoulder' on one side of the peak in some panels. The first moment average will therefore not coincide with the peak of the spectrum but will drag the average velocity towards zero. Even when the asymmetry of the spectrum is slight, the velocity deviations we need to detect are $<0.5$~km~s$^{-1}$. Therefore, assuming a symmetrical spectral profile introduces a significant error. In protoplanetary discs, the CO lines are optically thick so the emission traces the velocity structure at the $\tau \approx 1$ surface. Analytical descriptions of warped discs implement a thin disc approximation in which the angular momentum unit vector $\boldsymbol{\hat{L}}$ is constant in each cylindrical annulus. Due to the curve of the disc midplane, $\boldsymbol{\hat{L}} = \boldsymbol{\hat{L}} (r) $ and  $\boldsymbol{\hat{L}} $ for similar radii are not parallel (see Fig.~\ref{fig:diagram}). Therefore, the projected velocity $v_{\rm{los}} = v_{z\prime}(z\prime)$ (where $z\prime$ is the distance along the line of sight) and the depth of the emission surface in a thick disc becomes relevant. The question of whether the CO emission accurately traces the velocity structure in the disc therefore becomes more complex. For the purposes of comparing observations with simulations, the $v_{\rm{peak}}$ map matches the simulation projected velocities more closely because the peak is not reduced by the averaging. The higher values will be easier to detect and show up more structure. 
%
%sketch diagram
\subsection{Effect of optical depth}
\begin{figure*}
\centering
\includegraphics[width=\textwidth]{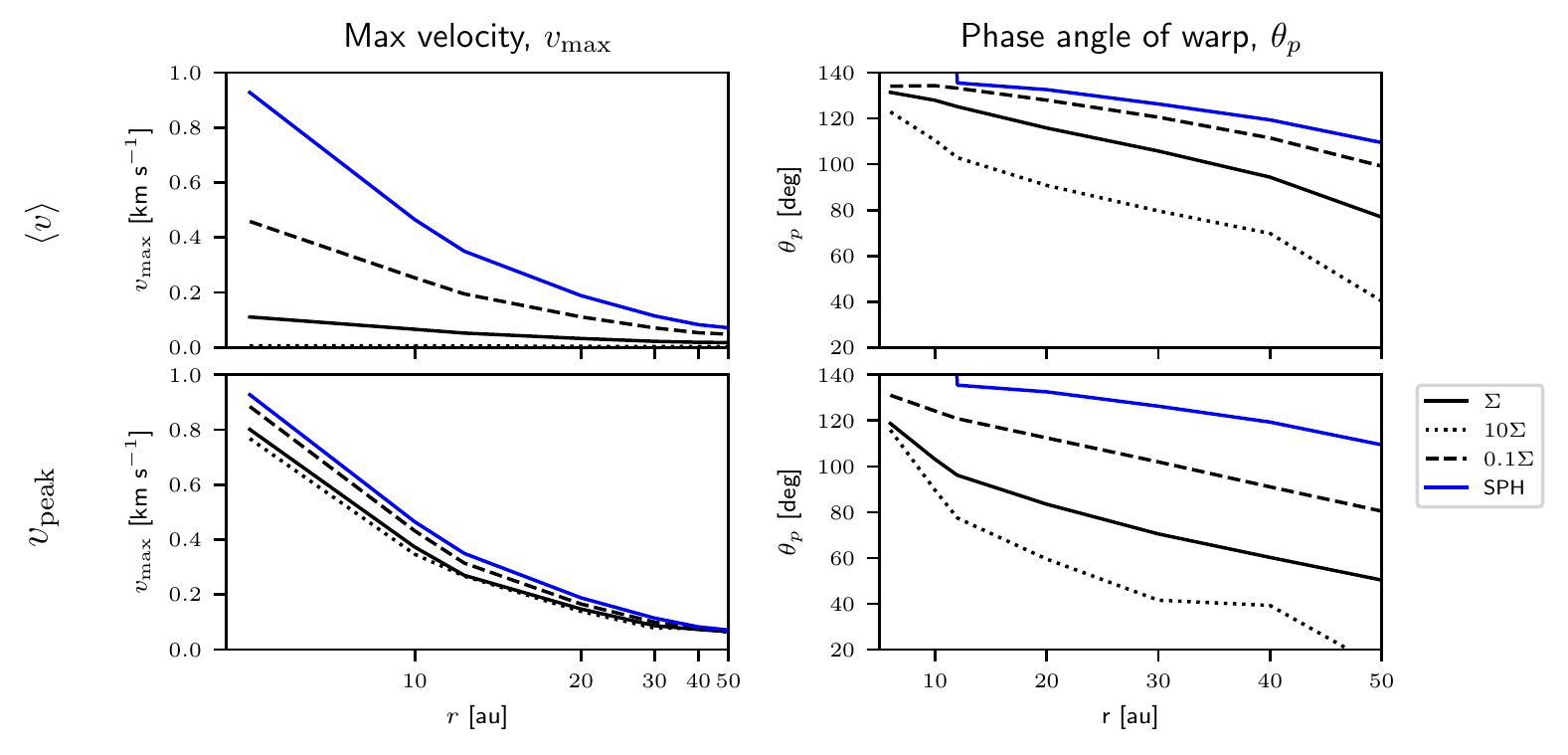}
\caption{Effect of optical depth/surface density. $v_{\mathrm{max}}$ and $\theta_p$ profiles as for Fig.~\ref{fig:sinefits} for the same disc model. The profiles were derived from simulated $^{13}$CO 3-2 line emission but with the surface density scaled by factors of 10 and 0.1 ($10\Sigma$ and $0.1\Sigma$).}
\label{fig:tauplots}
\end{figure*}

The projected velocity of a warped disc is not constant along the line of sight, as discussed earlier. Consequently, emission originating from different optical depths can be expected to give different values of projected velocity for the same spatial position. We simulate $^{13}$CO 3-2 emission maps scaling the disc surface density by factors of 0.1 and 10 in addition to the original value as a proxy for discs of differing the optical thickness. The resulting radial profiles of $v_{\mathrm{max}}$ and $\theta_p$ can be found in Fig.~\ref{fig:tauplots}.

The first moment maps, again, underestimate the velocity and phase angle in all cases. The lower the surface density, and the more optically thin the disc is, the closer the observationally derived values are to the those from the SPH simulation. This can be understood simply because a greater column of gas is contributing to the total emission due to the reduced effect of self-absorption. The SPH mass-weighted column-averaged velocity can be thought of as the perfectly optically-thin limit. The {\aky magnitude of the peak velocity} derived from the $v_{\rm{peak}}$ maps decreases with increasing opacity, {\aky although the values are closer to the simulation values than the values derived from first moment maps.}

The phase angle of the warp has a greater shift relative to the SPH model with increasing opacity. For both methods of deriving the velocity, the $10\Sigma$ model has a substantially greater change in $\theta_p$ with radius than the SPH model. This indicates that the opacity of a disc has a major contribution to the appearance of twisted velocity structures. This is an important result to consider because it means a distorted disc may appear more twisted than it really is if observed with an optically thick tracer.
%
%
%%%%% Inclination/deprojection.
\subsection{Detectability in an inclined disc}
\label{sec:results-inc}
\begin{figure*}
\centering
% made with inc_plots.py
\includegraphics[width=\textwidth]{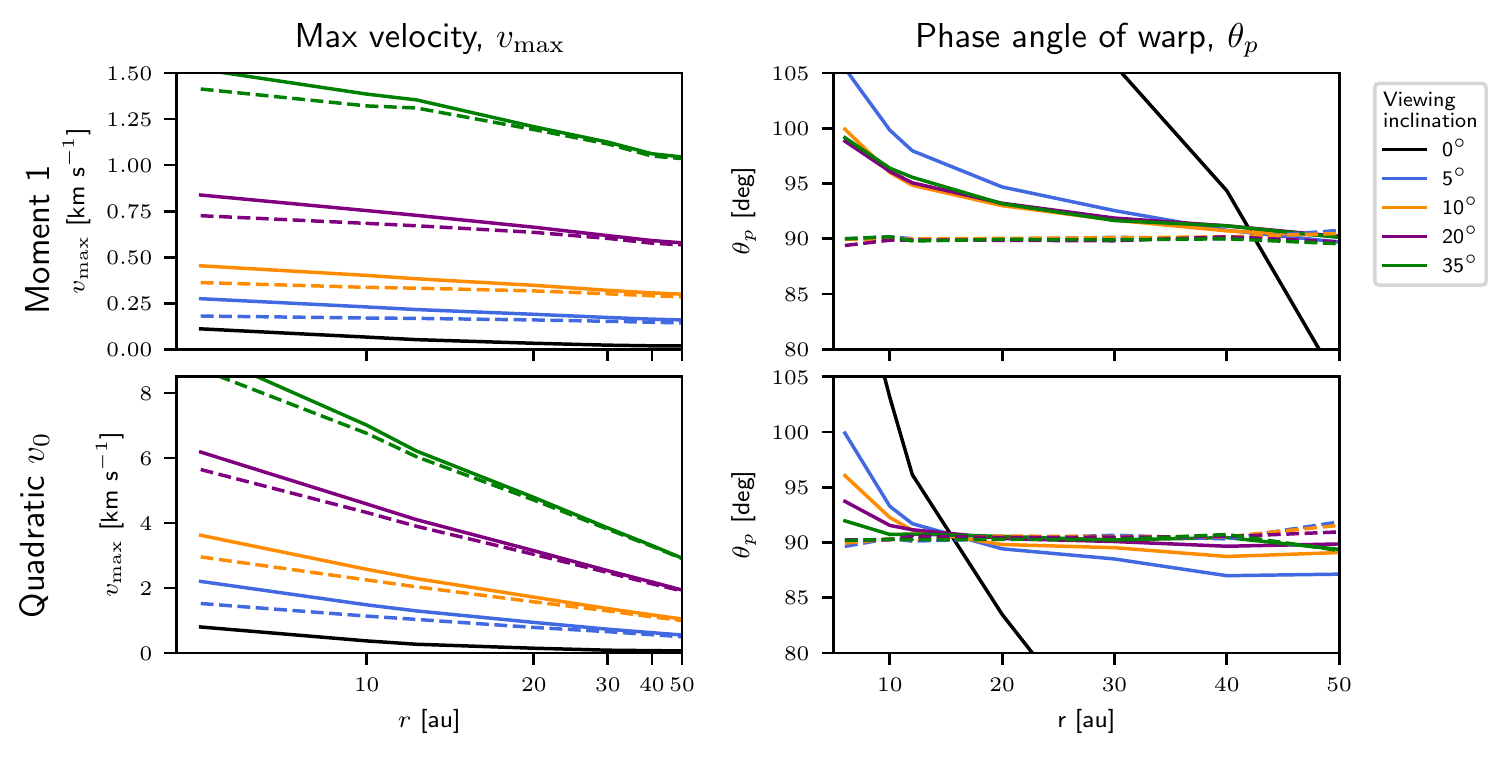}
\caption{The velocity and warp phase angle measurement when disc is viewed at an inclination for the $10^\circ$ binary inclination warped disc (solid lines) and the unwarped aligned disc (dashed lines). The velocity maps were spatially deprojected first but no Keplerian subtraction is applied. }
\label{fig:inclination}
\end{figure*}

The main challenge with detecting a small warp is that velocity deviations are likely to be smaller than the line of sight velocity due to the disc rotation at most inclinations. We extract the radial maximum velocity and warp phase angle profiles from spectral cubes produced from viewing the $10^\circ$ binary inclination warped disc at inclinations up to 35$^\circ$. The same analysis is repeated with the unperturbed aligned disc to look for differences due to the warp. The velocity maps were first deprojected spatially such that the disc appears circular but the velocity values were unchanged and not Keplerian subtracted.

The profiles are shown in Fig.~\ref{fig:inclination}. The differences in $v_{\mathrm{max}}$ between the warped and aligned discs are very small outside of 10 au with even a slight inclination. The change in $\theta_p$ with radius is far smaller when the disc is viewed at an inclination. This change is only a few degrees between 20-50~au but any deviation from a flat profile, i.e. a twist, is indicative of a warp. The detection of a warp in this manner is therefore only limited by the noise and spatial resolution of the data. The greatest deviations are seen where the warp is strongest, in this case at $r<20$~au, so the focus needs to be on observations suited to that scale for the best chance of detection.

\subsection{The effect of deprojection errors on warp characterisation}
\begin{figure*}
\centering
\includegraphics[width=\textwidth]{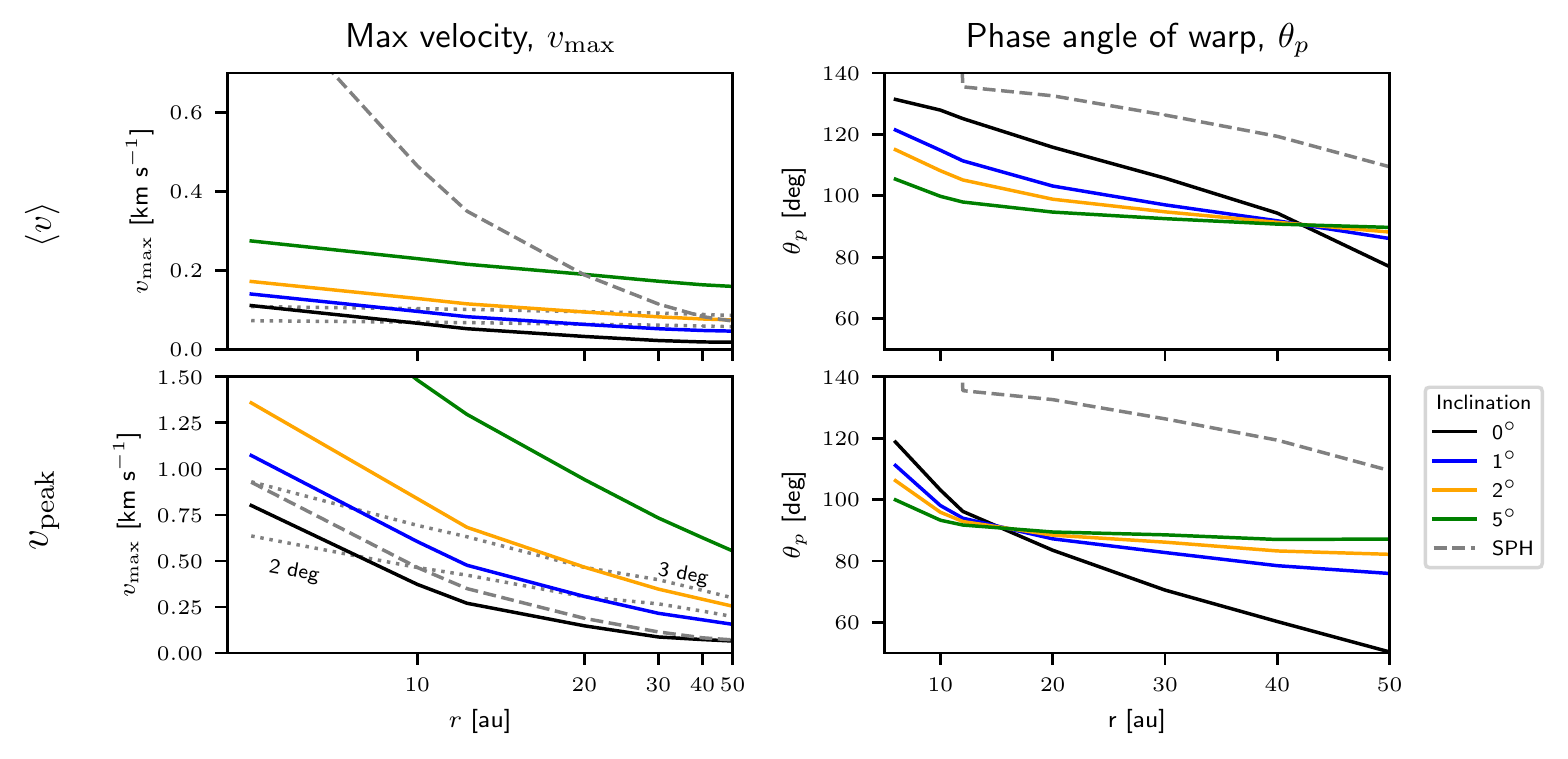} % made with sine_plots.py
  \caption{Fitted $v_{\rm{max}}$ and $\theta_p$ profiles from inclined velocity maps, equivalent to small errors in the inclination used for deprojection. Note the different vertical scales in the left column. The labeled dotted lines show the velocity profiles for the aligned circumbinary disc viewed at 2 and 3\degree ~for comparison.}
 \label{fig:sinefits}
\end{figure*}
As observational methods are increasingly capable of measuring velocities to $\sim 30$~m~s$^{-1}$ precision, the derived gas velocities are susceptible to the effects of systematic errors in inclination and position angle which are generally only known to the nearest $\sim 5 ^\circ$. {\aky These errors will introduce an additional velocity field in the same way as an additional inclination.} Fig.~\ref{fig:sinefits} includes the radial $v_z$ and $\theta_p$ profiles for viewing inclinations 1, 2, and 5$^\circ$, which is equivalent to an error in deprojection by the same value (in $i$, assuming the position angle is accurate).

The velocity profile for a circumbinary disc in which the orbital planes of the binary and disc are aligned are plotted for comparison in Fig.~\ref{fig:sinefits} for a small viewing inclination. When viewed face-on, the velocity profile of the warped disc differs most from the linear profile of the aligned disc but this flattens off with increasing inclination. With a $5^\circ$ error in inclination, a warp of this amplitude is likely to be missed. The greatest deviations are seen in the inner 20~au in $v_{\rm{peak}}$, with velocities exceeding 0.5~km~s$^{-1}$. Pinning down the velocities in the inner regions of the disc where the warp amplitude is greater is therefore important so it may be worthwhile compromising the spectral resolution slightly to achieve greater spatial resolution.

The presence of a warp is most clearly seen in the `twist', or variation of $\theta_p$ with radius (Fig.~\ref{fig:sinefits}). At a viewing inclination $i=5^\circ$, for $20 < r<50$~au the change $\Delta\theta_p$ is $\sim 5^\circ$ which is likely to come close to the uncertainty due to noise in the data. If radii down to 10 au are available, $\Delta\theta_p$ increases to nearly 15$^\circ$. If the section of the disc where the warp amplitude is greatest is observed, then small warps should be readily detectable if the error in deprojection is no greater than $\sim 5^\circ$.

\begin{figure*}
\centering
\includegraphics[width=\textwidth,trim= 5cm 8cm 4cm 3cm ,clip]{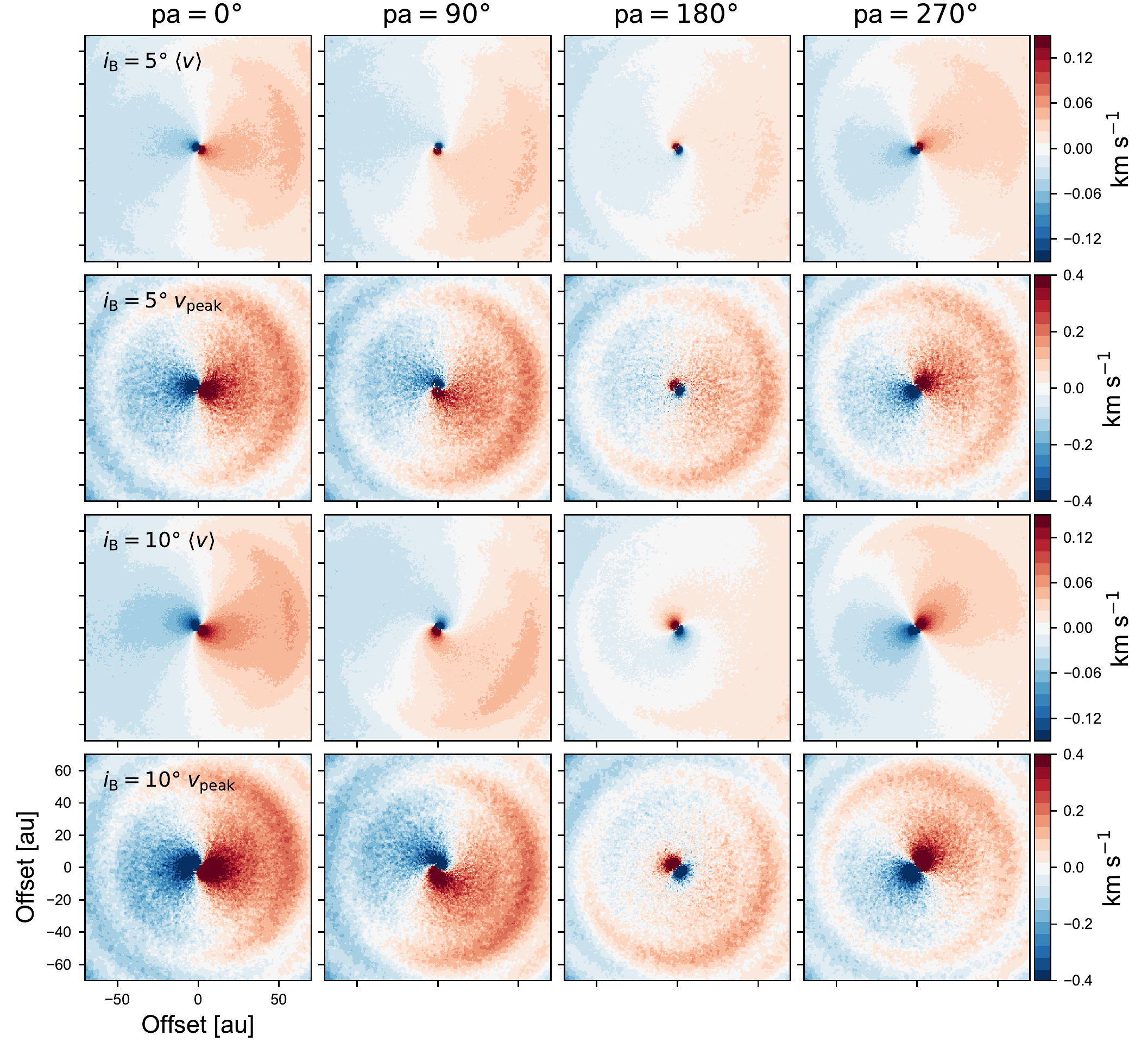}
\caption{The effect of warp position in the disc. $^{13}$CO first moment and $v_{\rm{peak}}$ maps for the $5^\circ$ and $10^\circ$ binary inclination models viewed at an inclination of $1^\circ$ with the warp rotated to four different position angles. The clarity of the `twist' is variable becoming barely discernible in some cases.}
\label{fig:CB5CB10}
\end{figure*}

\begin{figure*}
% made with sin_plots_PA.py
\centering
\includegraphics[width=\textwidth]{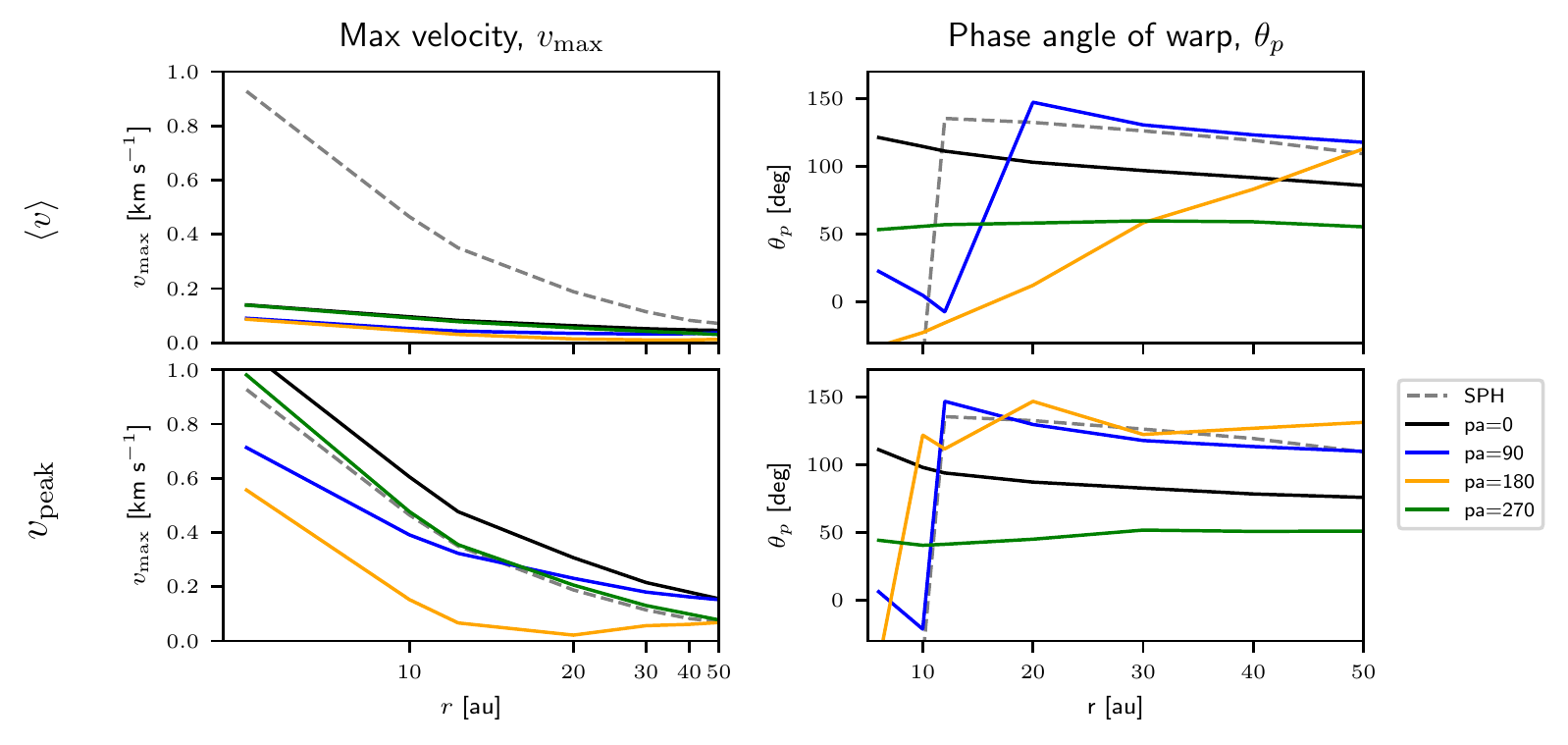}
\caption{Dependence on position angle of warp in disc. $v_{\mathrm{max}}$ and $\theta_p$ profiles as for Fig.~\ref{fig:sinefits} for the same disc model. The profiles were derived from simulated $^{13}$CO 3-2 line emission for a viewing inclination of $1^\circ$ with the disc first rotated such that the warp is at a different position angle.}
\label{fig:PAplots}
\end{figure*}

%*** Position angle*** 
The appearance of the warp is affected by its azimuthal position in the disc. In Fig.~\ref{fig:CB5CB10}, we demonstrate the appearance of a warp with just a $1^\circ$ error in the inclination used for deprojection at four azimuthal positions relative to the assumed line of nodes. Just by eye, there is an apparent difference in both the velocity profile and the twist. Indeed, the radial profiles in Fig.~\ref{fig:PAplots} for $i=1^\circ$ confirm this. For a 270$^\circ$ position, the twist produced by the warp is imperceptible. For a more detailed study of this effect, the reader is referred to \citet{juhasz2017}.
\subsection{Channel maps}
\begin{figure}
%\centering
  \includegraphics[width=\columnwidth,trim= 0.1cm 0.6cm 0.95cm 1.2cm, clip]{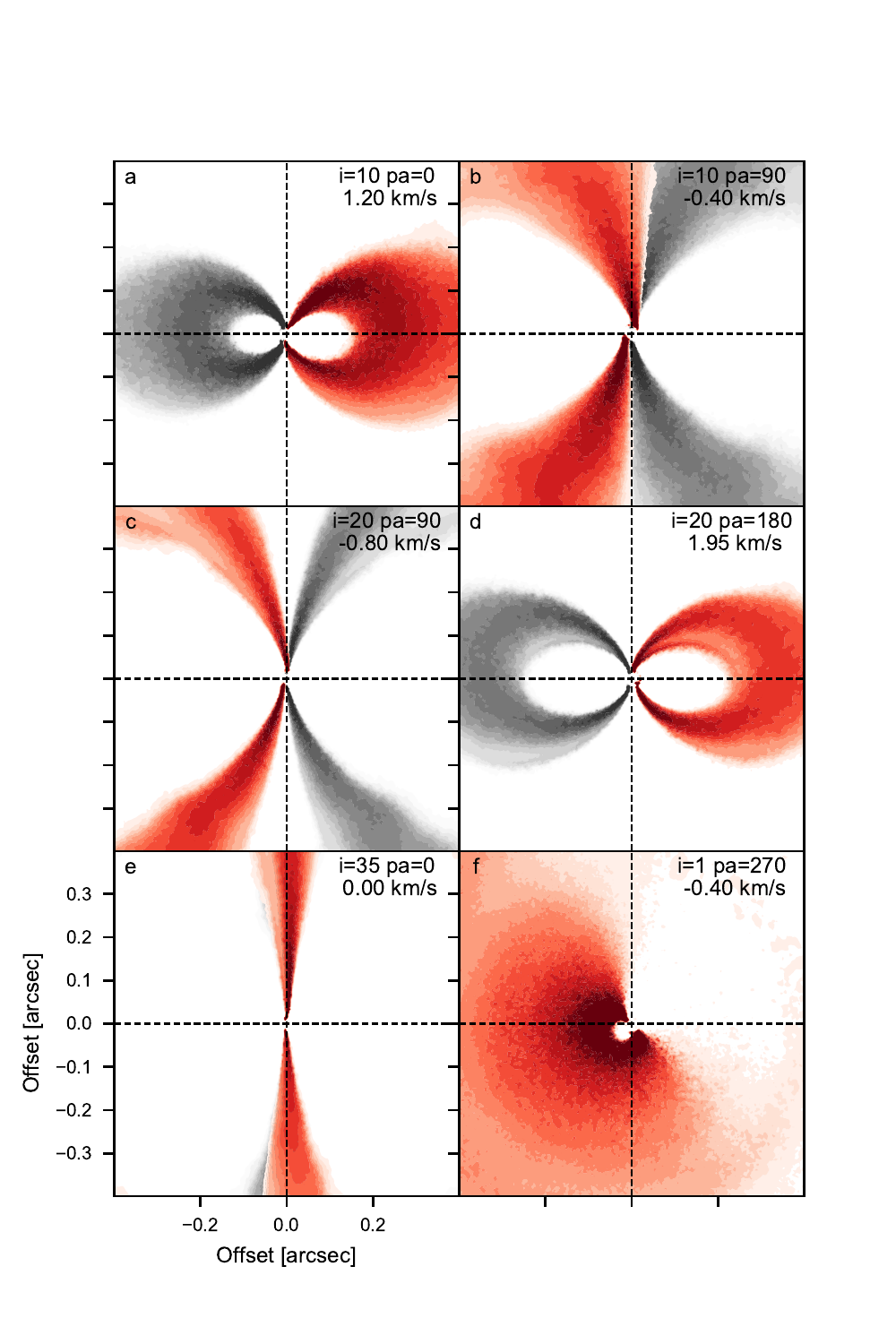}
  \centering 
  \includegraphics[width=0.5\columnwidth]{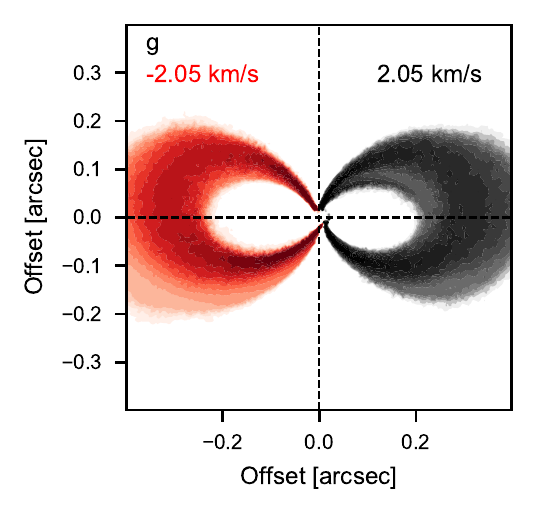}
    \caption{Continuum-subtracted channel maps for the $10^\circ$ binary inclination model at various viewing inclinations and position angles (red shading). The corresponding channel maps of the aligned disc model are shown in grey (except panel f), reflected along the vertical axis for comparison to highlight the asymmetries in the emission from the warped disc. Panel g shows two channel maps from i=20, pa=180. Here we see asymmetry between opposite velocity channels. The \aky{back side of the disc is just detected at positive velocities (red) but not at negative velocities (grey.}}
   \label{fig:6Kchanmaps}
\end{figure}

Examining the channel maps can reveal small deviations from a smooth Keplerian rotation profile that would otherwise have gone undetected. We find asymmetries in the emission across several channels. In Fig.~\ref{fig:6Kchanmaps}, selected channel maps produced for the $10^\circ$ binary inclination model viewed at various inclinations are presented alongside the corresponding channel from the aligned disc model. In Fig.~\ref{fig:6Kchanmaps} (a) the red-shifted emission crosses over onto the opposite side of the disc at the centre. Similarly, in panel (b) the blue-shifted emission crosses to the opposite side, but only for the far side of the disc. Panel (c) shows a brightness asymmetry and the emission from the near and far side is offset from each other. In panel (d) emission from the lower surface of the disc is faintly visible in the warped disc but not in the aligned disc. Panel (e) shows that, even at a moderate inclination, a difference in offset of the emission of a central velocity channel may be detectable between opposite sides of the disc. Emission from the reverse side of the disc is faintly visible towards the top of the image. The channel map in panel (f) is from the velocity cube for which the $v_{\rm{peak}}$ map in Fig.~\ref{fig:CB5CB10} and the change in $\theta_p$ (Fig.~\ref{fig:PAplots}) showed little evidence of a twist. Nevertheless, the channel map is asymmetrical, revealing the warp. Panel (g) shows the asymmetries between {\aky positive and negative} velocity channels for the same disc.

These asymmetries are seen for small warps even at a moderate viewing inclination and therefore provides a method for detecting small warps when it is not possible to discern them from the velocity maps due to the viewing inclination. In addition, it is likely that this approach will not require as high velocity resolution because the effect is not limited to a narrow range of channels. The asymmetries extend over tens of au, equivalent to multiple synthesized beam widths with ALMA. Consequently, the observed structures should be robust against noise.

\begin{figure*}
\centering
\includegraphics[width=\textwidth]{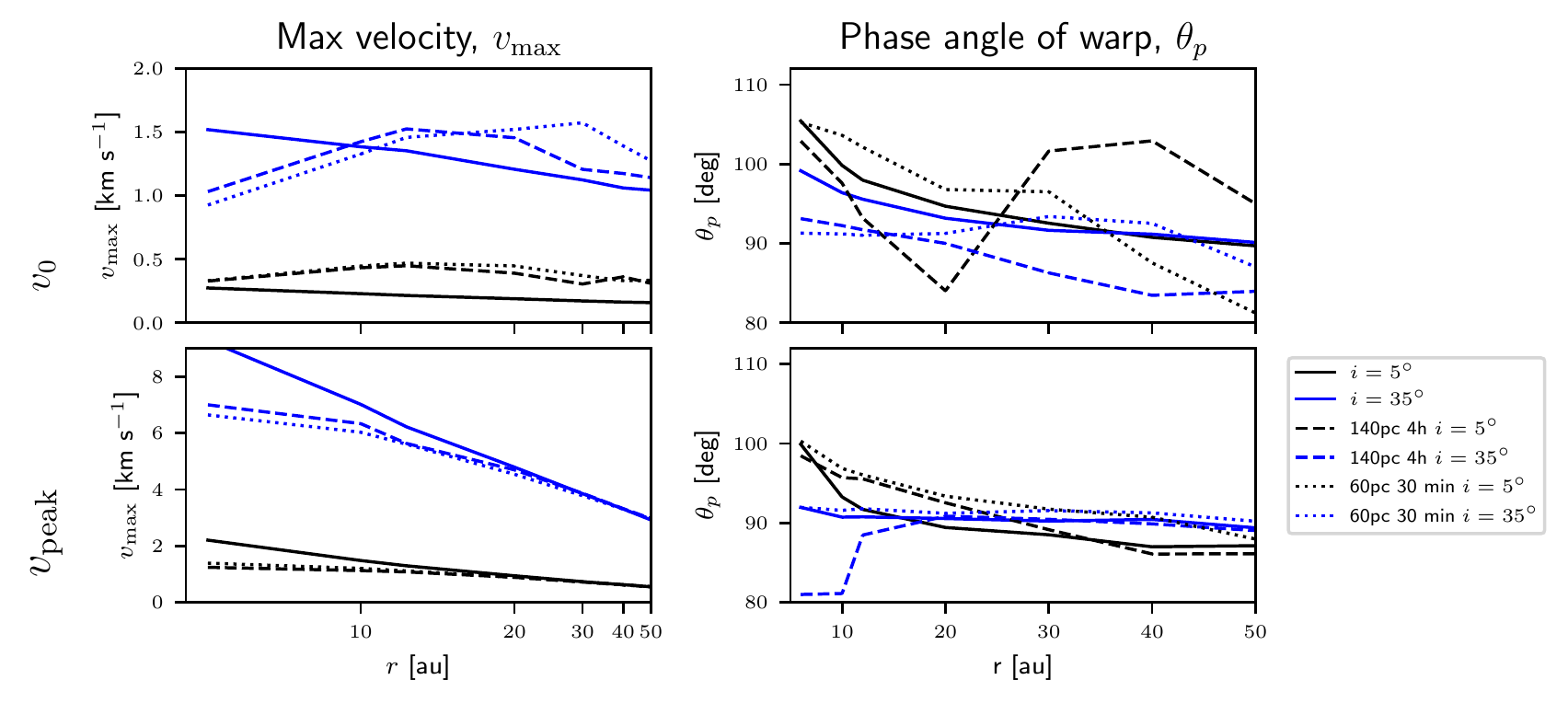} % made with sine_plots_noise.py
  \caption{\aky A comparison of the derived velocity and phase angle profiles from the synthesised CO data with added thermal noise and convolution with a Gaussian beam of 0.05 arcsec. The thermal noise was chosen to match the sensitivity of integration times of 4 hours and 30 minutes with the full ALMA array and we show the profiles from a distance of 140pc and from fluxes scaled to a distance of 60pc.}
 \label{fig:noisysinefits}
\end{figure*}

\subsection{The effects of noise and spatial resolution}
\label{sec:noise}
{\aky
The results we have presented so far are derived from idealised noise-free synthetic observations. We now demonstrate that the observational characteristics of small warps that we described are recoverable with a representative spatial resolution and signal to noise ratio. The issues of image fidelity when imaging interferometric data are known difficulties to observers. The methods used to mitigate these problems will depend on a number of factors to do with the specific observation. For this reason, we consider a generic case as an example. We add thermal noise to the data and convolve with an 0.1 arcsec Gaussian beam to emulate representative quality of data. We compare the results obtained from the synthetic line emission calculated for a distance of 140~pc and from the same synthetic data but with the flux scaled for a distance of 60~pc, the approximate distance to TW Hydrae. At 60~pc, the flux density in the warped region of the disc is $\sim 220$~mJy~beam$^{-1}$ and at 140~pc is $\sim 40$~mJy~beam$^{-1}$, where the beam size is 1 arcsec. Gaussian noise was added with $\sigma = 6.2$~mJy and $\sigma =17$~mJy, representing the sensitivity of the ALMA 12~m array at a spectral resolution of 0.05~km~s$^{-1}$ for a 4~hour and 30 minute integration respectively. In creating the velocity maps with {\sc bettermoments}, the velocity was smoothed over 5 channels. The velocity maps were calculated for between $\pm 4.0$~km~s$^{-1}$ for $i=5^\circ$ and between $\pm10.0$~km~s$^{-1}$ for $i=35^\circ$ to reduce the effect of noise.

The derived velocity and twist profiles from the convolved data for the 10$^\circ$ binary inclination at viewing inclinations of $5^\circ$ and $35^\circ$ are compared with the idealised results presented earlier in Fig.~\ref{fig:noisysinefits}. At a distance of 140~pc, the beam size is 14~au which means that the velocity structure is not resolved in the inner regions and the derived values from the convolved velocity maps deviate from the idealised model. At $r>20$~au, where the disc is adequately resolved, the $v_{\rm {peak}}$-derived velocity profile follows the idealised model closely. The same is true of the twist profile at $i=35^\circ$ but at $i=5^\circ$ we see an increased twist. The velocity and twist derived from the first moment maps suffer most from noise and poorer resolution. From this, it is clear that better quality data is obtained from the peak velocity maps. 

For the both distances and integration times, the maximum velocity profile is recovered reasonably well from the $v_{\rm {peak}}$ maps where the disc is resolved. In contrast, the phase angle profile is sensitive to noise and the finite resolution. The width of the annuli chosen for extracting the phase angle is a trade-off between averaging out noise and being narrow enough to resolve the change in phase angle with radius, noting that the twist cannot be extracted on scales smaller than the beam size. Given these issues, the methods of analysis presented here are more suited to nearer protoplanetary discs, unless a more distant disc happens to be particularly extended and bright in CO. In the scenario presented here, the twist does not show up in the $v_{\rm {peak}}$ map at $i=35^\circ$ even for the 60~pc disc, indicating that lack of evidence of a twist does not necessarily discount the presence of a warp.}

\section{Discussion}
\label{sec:discussion}
\subsection{Comparisons to related work}
Identifying small warps kinematically is naturally challenging because the velocity perturbation is small. Nevertheless, we have identified three kinematic signatures of slightly warped protoplanetary discs. The radial profile of the projected velocity deviates from the expected near-Keplerian profile, becoming much steeper in the warped region. By fitting a sinusoid to the azimuthal velocity profiles for concentric annuli we can quantify the twist and this will facilitate direct comparisons between discs. Lastly, individual channel maps present various asymmetries and distortions, even at a moderate inclination of 35$^\circ$. Depending on the viewing inclination, not all of these effects may be observed.

{\aky Previous theoretical work to determine the observability of non-axisymmetric structures such as circumplanetary discs and planet-induced wakes usually implements a fitting of a rotation curve to the whole disc (e.g. \citealt{rosenfeld2013b,pinte2018,perezS2018,teague2019}). This method works well when there is a consistent inclination throughout the disc but for a warped disc, the velocity field is twisted and the projected radial velocity profile deviates from that of a Keplerian disc. The position angle and inclination vary with radius which means that these parameters cannot be consistently fit for the whole disc. For the purposes of deprojection, it might be better to estimate these values for the outer disc from continuum observations. \citet{casassus2019a} implement a method which accounts for radial variation in inclination and position angle and also fits for the height of the emitting layer in the disc. In this paper, we also fit annuli extracted from the disc rather than the whole disc. This approach is better suited for warped discs, since pixels in a given annulus correspond to regions of the disc with similar orientations. We do not look to measure the absolute orientation of the annuli, but rather the differences between them to obtain a quantitative measure of the twist.

Resolved kinematic observations or protoplanetary discs typically consider the height of the emitting gas layer (e.g. \citealt{pinte2018a,casassus2019a}) to map the rotation profile more accurately, accounting for the fact that protoplanetary discs are not razor thin. The height of the $\tau=1$ surface in a warped disc is a complex shape. Any analytical $z/r$ profile would not be accurate and would not improve the accuracy of the results so we do not consider this here.}

\subsection{Sources of uncertainty}
%measuring velocity
Ideally, we would like to measure the velocity structure of a warped disc. The out-of-plane gas motions are easily detected when the disc is viewed face-on, but are swamped by the rotation and quickly become indistinguishable with just a few degrees inclination. An important question is how closely the observed velocity field follows the source velocity field. We have shown that the velocity field obtained from first moment maps differs significantly from the source gas velocities. This is because the average obtained from the first moment is only valid if the spectrum is symmetrical, which is not true for a warped disc. The quadratic method of calculating $v_{\rm{peak}}$ of \citet{teague2018} recovers the model structure more accurately but there will be some dependence on the optical depth of the tracer.

%error in deprojection - how to measure i and pa, Eddy fitting
The methods employed above for deriving the radial profiles of projected velocity and phase angle of the warp (twist) assume that the deprojection of the disc observation is perfect. The inclination and position angle is difficult to constrain and is rarely constrained to better than a few degrees accuracy. The most accurate way of estimating these values is by fitting Keplerian disc models to the observations. This is a problem when we are dealing with a disc that is not flat. There have been attempts to produce analytical models of warped discs for the purpose of fitting observations, however this introduces more free parameters to describe the shape of the warp and the analytical disc model does not always replicate the morphology seen in hydrodynamical simulations. Brief testing  of fitting simulated CO observations with {\sc eddy} (\citealt{teague-eddy}, see Appendix~\ref{sec:eddyfitting}) showed that no good fit could be obtained for warped discs that were close to face on and for inclined warped discs the central mass tended to be overestimated or the fitted centre of the disc was offset from the true centre. The resulting model-subtracted images often showed spiral features, presumably a product of the poor fitting. The fitting was improved by masking out the warped region but this naturally requires some prior knowledge and the results can be sensitive to the selected region. If there is a clear twist in the inner regions and a large enough near-Keplerian region of the disc beyond, this method may perform more accurately. It may also be possible to mask out channels close to zero since for lower velocities the perturbation due to a warp is a larger fraction of total. When the velocity map appears non-Keplerian, this kind of fitting is naturally unsuitable and we must rely upon continuum imaging to estimate the inclination and position angle.

%uncertainties arising from deprojection errors
We expect that the disc inclination may only be known to within 5$^\circ$ and have shown that this error in deprojection can conceal the kinematic signatures of small warps. The error serves to flatten the projected velocity profile which makes it less distinct from that of a smooth near-Keplerian disc and also reduces the observed twist. For small warps, the deprojection needs to be accurate to $\sim5^\circ$ or better to reliably pick up the kinematics, so this uncertainty is a major limitation. We have shown that velocity perturbations caused by small warps can be discerned in the continuum-subtracted channel maps, which avoids the need for deprojection.

Supposing the orientation of the warp in the disc and the disc on the sky are well constrained, we estimate that the velocity measured from the $^{13}$CO $v_{\rm{peak}}$ map is within 0.1 km/s or around 25 per cent of the true mean gas velocity for the case studied here. Differences in the optical depth are expected to have a smaller effect on the observed velocity since we saw a change in the $v_{\rm{peak}}$ profile of $>0.1$~km~s$^{-1}$ with a factor 100 change in the disc surface density. The disc appears more twisted with increased opacity which means we cannot take the warp morphology at face value. An optically thin tracer is necessary to recover the true morphology of the warp. Before we can consider linking the observed disc structure to its physical properties and those of the perturbers, we need to be aware of these limitations and of the resulting uncertainties or find ways to mitigate them.

{\edit At moderate inclinations, any radial flows will contribute to the radial velocity and twist profiles. The twist due to the warp has a greater phase angle at low inclinations. The component of the projected velocity that is due to radial flows increases with inclination. This means that at near-face on inclinations, any radial flows would have a negligible contribution to the radial velocity and twist profiles, compared to that of the warp. At moderate inclinations, however, the possible presence of radial flows is an additional caveat.}

{\aky We showed in section \ref{sec:noise} that a representative spatial resolution and thermal noise can change the observed twist at low inclinations. Crucially, there is still a detectable change in phase angle and the maximum velocity profile derived from the convolved $v_{\mathrm{peak}}$ map reproduces the idealised profile where the disc is resolved, despite the noise and reduced resolution.
}
%
%
 %%%%%%%%%%%%%
\subsection{Are we missing warps or misinterpreting them?}
%%%%%%%%%%%%%
%
{\aky The prevalence of broad shadows in scattered light images of protoplanetary suggests that misalignments are common but a pronounced twist in the velocity field is currently the primary method of identifying a warp.} Additionally, given the prevalence of binary and higher order multiple systems, as well as the somewhat chaotic nature of the accretion onto protostellar discs, for example in the models of \citet{bate2018}, we would expect warps to be common. This raises the question of why more warped discs have not been identified. Perhaps we are missing them due to the methods used to analyse kinematic data or perhaps the signatures of warps are misinterpreted or discounted.

As discussed earlier, many methods of analysing disc kinematics assume that the spectrum is Gaussian, or at least symmetrical. This conceals the velocity deviations due to the warp and gives a misleading velocity field in the first moment map. Attempts to fit a Keplerian disc model will give incorrect results. This may lead to a warp going unnoticed. If a Keplerian model is fitted and subtracted, the resulting residuals may show spurious spiral arm features rather than an obvious warp.

The velocity deviation is a small fraction of the rotation velocity so will be tricky to detect above this background. We base most of the analysis here on the deprojected observed disc and have examined how the derived warp properties are affected by small errors in the inclination. If we can observe the region of the disc that is warped at high angular resolution then the maximum velocity profile can still be observed to be steeper than that of an unwarped disc. In the model here, the region were the warping is strongest is within $r<20$~au and the velocity and twist profile are likely to be difficult to distinguish from that of an unwarped disc outside this region, depending on the noise and uncertainty in the maximum value of the velocities. The scales explored here is determined by the binary separation of 1~au. Small warps with larger angular size will of course be easier to resolve spatially. The implication here is that a warp could well be missed if observed at a spatial resolution too low to resolve the warped region. The asymmetries in the channel maps are small so it is possible that they would be missed if one is not looking for them. If there is no obvious twist in the observed velocity field, perhaps until now there would be no reason to look for evidence in the channel maps.

Depending on the azimuthal position of the warp, the twist may not be evident. We have seen, as also found by \citet{juhasz2017} and \citet{facchini2018}, that the relative positions of the line of nodes and the phase angle of the warp maximum determines whether the maximum perturbation is visible or not, and consequently determines the observed velocity profile. In these cases, the perturbation can still be seen in the channel maps.

{\aky{\edit Radial infall within a cavity gives similar twist and channel maps. Complementary submillimetre continuum imaging will confirm whether there is a central cavity and scattered light images will reveal shadows due to misalignments. Currently, these are the only features that may distinguish the two scenarios. 
}}

Another possibility is that small warps are being misinterpreted as other structures. Embedded planets can also cause velocity deviations (e.g. \citealt{perezS2015,teague2018jun,pinte2018}). These are, however, very much localised, occurring in just one velocity channel and confined to a small region of the disc near the planet. The observed velocity field of a warped disc is also distinct from that of a gravitationally unstable disc with a spiral structure. In such discs we expect to see a finger-like structure in the first moment map protruding from the red-shifted side into the blue-shifted side of the disc \citep{hall2020} and the velocity perturbations are coincident with spiral arms. In a warped disc, there may be finger-like perturbations in the velocity map, outside of the central twisted region, but these are not coincident with spiral arms in the gas density.

The vertical shear instability can produce ring structures in line-of-sight velocity maps \citep{barraza2021} with a wiggle-like effect around the disc's line of nodes which bear some resemblance to the velocity maps we present for warped discs and have a similar magnitude. The key difference is that a warped disc has a twisted inner region whereas the vertical shear instability leaves a symmetrical velocity field in the inner disc and axisymmetric rings in the first moment map.

When there is a planet embedded in the disc, other perturbations can occur such as meridional flows (e.g. \citealt{teague2019Nat,casassus2021}). Care should be taken for regions that are close to a gap in the disc or a cavity because additional perturbations may be contributing to the velocity field.

\subsection{\aky Possible shallow warped discs from the literature}

The effect of a small warp on the kinematic observations is subtle and we propose that many protoplanetary discs may be warped but the evidence has not been examined fully. Here we compare the predictions for the characteristic kinematic features of warped discs with three protoplanetary discs from the literature. {\aky As discussed earlier, the warp in our simulations is caused by a misaligned central binary but the structure is similar for warps resulting from various effects so the effects seen in line emission are also similar.}
\subsubsection{TW Hydrae} 
{\aky The suggestion that the centre of the TW Hya disc is warped originated from azimuthal variations in the scattered light surface brightness \citep{roberge2005}. There are several published CO observations of TW Hya and we now compare these with the results of the modelling.}
\citet{rosenfeld2012} studied the CO line wings in the inner $\sim4$~au of the TW Hya disc. They describe an excess of emission in the line wings which is fitted well by a model including a warped inner disc. {\aky However, the synthesized beam was $> 90$~au, precluding further comparison and the higher resolution channel maps published by \citet{huang2018a} show no obvious twist. \citet{debes2017} and \citet{poteet2018} measured the rotation of the optical surface brightness asymmetry from observations at 4 epochs. The rotation speed of the asymmetry led to the conclusion that there is an inner disc or warp within $r<1$~au precessing on the timescale of 17 years. Such a feature is below the resolution of the CO observations described above. \citet{teague2018} tease out very small variations in the velocity field and CO brightness temperature, revealing a spiral. In the velocity residuals (their Fig. 4) there is an `X' shape at $r<0.2$~arcsec and this is indicative of non-Keplerian motion and consistent with a compact warp. Apart from the spiral, there is no further asymmetry discernible. If the shadowing were to affect the temperature in the disc, we would expect to see asymmetry in the molecular line emission. However, this effect is unlikely to be seen in the $^{12}$CO line which originates in the lower density surface region of the disc \citep{casassus2019} and an analysis of $^{13}$CO may be more informative. Molecules whose abundances are driven by radiative rather than thermal processes, such as HCO$^+$, may still present asymmetric emission \citep{young2021}. \citet{oberg2021} present DCO$^+$ observations in TW Hya. While there is no obvious asymmetry in the images, it would be worth searching for any small azimuthal variation that would be colocated with the location of the shadow at the time of observation. In summary, TW Hya is probably an example of a disc with a warp that is extremely difficult to detect because the angular scale of the warp is small and the misalignment is only very slight. In addition, the dust is depleted at the location of the possible warp or inner disc, which will further reduce the effect of shadowing. }
\subsubsection{HD~163296}
In the simulated channel maps of a disc with a small warp, we see that the emission from opposite sides of the disc is slightly offset, especially in the central velocity channels, rather than being symmetrical as is the case for the aligned disc (see Fig.~\ref{fig:6Kchanmaps}). The opposite sides of the disc in the channel maps of HD~163296 presented in \citet{pinte2020} appear to be similarly offset from one another. This is separate to the `kinks' identified by the authors which were attributed to planets. Previously, time-varying asymmetry has been observed in scattered light images \citep{rich2019}, indicating time-variable illumination of the outer disc. A precessing inner disc and/or warp was posited to explain this. The larger scale asymmetry in the channel map would support the warp hypothesis. There may additionally be a very small precessing inner disc, like has been proposed for TW Hya, and this too would be accompanied by a warp.
\subsubsection{Elias 2-27}
The situation may be complicated when multiple effects are present. \citet{paneque2021} suggest that the Elias 2-27 disc is warped to explain the observed azimuthal asymmetry in the gas emission but the disc also hosts gravitational instability (GI) induced spirals \citep{perez2016aa}. \citet{veronesi2021} also find Elias 2-27 to be susceptible to GI. The $^{13}$CO velocity map presented by \citet{paneque2021} shows the same slight twist as in some of the velocity maps in Fig.~\ref{fig:CB5CB10}. The residuals for the $^{13}$CO and C$^{18}$O first moment maps of Elias 2-27 and the best fit model indicate a non-Keplerian velocity field. In appendix \ref{sec:eddyfitting}, we examine the expected residuals from attempting to fit a Keplerian model to a warped disc further. We find the same features in the residuals that \citet{paneque2021} obtained from observations of Elias 2-27 (see Fig.~\ref{fig:eddyfit}), which strongly suggests this disc is warped. Late-stage infall onto the disc may result in both GI and warping so we may expect these two effects to occur simultaneously. 
\section{Conclusion}
Many protoplanetary show signs of small misalignments through the shadows and asymmetric illumination seen in scattered light imagery. Through hydrodynamical and radiative transfer modelling of discs around misaligned binary systems, we have studied the effects of a small warp on the observed kinematics.

The key implications for identifying small warps in protoplanetary discs are as follows:
\begin{enumerate}
\item The line profiles of warped discs are asymmetrical and velocities derived from observations are different to the intrinsic gas velocities. The quadratic method of determining the peak velocity field \citep{teague2018} yields values closer to the intrinsic velocity than the standard first moment map. {\aky When noise is considered, this method still performs well, as long as the disc is resolved.}
\item Keplerian fitting and subtraction does not necessarily produce an accurate velocity field. At worst, it produces spurious spiral arms and central features in the residuals when constant inclination and position angle are assumed across the whole disc.
\item The degree of observed warping can be quantified by extracting concentric annuli from the velocity map and fitting a sinusoid to each annulus. The `twist' is then quantified as the change in position angle of the velocity peak with radius. This allows the detection of very small twists. {\edit Twists can also result from radial flows but for this model with a small warp the radial component of the gas velocity is $\lesssim 10^{-4}v_{\rm Kep }$, which causes a shift in position angle $\ll 1^\circ$ and has a negligible contribution to the twist in this case. }
\item The greatest deviations are seen where the warp is strongest, in this case at $r<20$~au. Observations must be suited to that scale for the best chance of detection. For example, the spatial resolution should be prioritised over spectral resolution in the case of a very compact warped region.
\item Imperfect deprojection reduces the observed `twist' and flattens the line-of-sight velocity radial profile making it more similar to that of an inclined, unwarped disc. The deprojection must be accurate to $\lesssim5^\circ$ to detect a small warp outside of the most strongly warped region through sinusoid fitting.
\item The velocity profiles of warped discs that are not close to face on are very difficult to distinguish from that of an unwarped disc {\edit since the contribution from the projected Keplerian motion dominates the velocity profile. The radius at which the warped disc velocity profile and the purely Keplerian profile become indistinguishable depends on the amplitude of the warp}.
\item Like strongly warped discs, discs with small warps display the same twist in velocity maps. The observed twist depends on the optical depth and does not necessarily correspond to the true disc structure. Optically thin tracers provide better agreement with the disc structure.
\item At moderate viewing inclinations, the warp causes {\edit asymmetries in the CO channel maps out to 50~au, beyond where the amplitude of the warp is at a maximum}. These features are distinct from those due to embedded planets (`kinks') and gravitational instability-induced spirals (`wiggles'). These effects were observed in models with a viewing inclination $i\lesssim 35^\circ$ {\edit (Fig.~\ref{fig:6Kchanmaps}, panel e.)}. The shift in the location of the emission can be subtle and may have been missed in previous observations.
\end{enumerate}
%
%%%% Last paragraph
The kinematic detection of small warps is tricky but possible in most scenarios. Kinematic observations provide additional evidence of warping where a warp is suspected from scattered light observations or an alternative method when a source is too embedded for useful infra red observations. At low inclinations, CO observations can provide insight into the morphology of the warp and at moderate inclinations the warp causes detectable asymmetries in the channel maps. This provides a step towards the quantitative comparisons of observed warp structures with analytical and numerical models.

%%%%%%%%%%%%

\section*{Acknowledgements}
We thank the reviewer, Simon Casassus, for his comments and suggestions that helped to improve the manuscript. AKY thanks Chris Nixon and Anagha Raj for useful discussions. This research made use of the DiRAC Data Intensive service at Leicester, operated by the University of Leicester IT Services, which forms part of the Science and Technology Facilities Council (STFC) DiRAC HPC Facility (www.dirac.ac.uk). The equipment was funded by BEIS capital funding via STFC capital grants ST/K000373/1 and ST/R002363/1 and STFC DiRAC Operations grant ST/R001014/1. DiRAC is part of the National e-Infrastructure. AKY and RA gratefully acknowledge funding from the European Research Council (ERC) under the European Union's Horizon 2020 research and innovation programme (grant agreement No 681601). AKY is grateful for support from the UK STFC via grant ST/V000594/1. G.R. acknowledges support from the Netherlands Organisation for Scientific Research (NWO, program number 016.Veni.192.233) and from an STFC Ernest Rutherford Fellowship (grant number ST/T003855/1). This work made use of {\sc scipy} \citet{2020SciPy}, {\sc numpy} \citet{harris2020}, {\sc matplotlib} \citep{hunter2007}, {\sc astropy} \citep{astropy:2013,astropy:2018} and {\sc SPLASH} \citep{price2007}. For the purpose of open access, the author has applied a Creative Commons Attribution (CC BY) licence to any Author Accepted Manuscript version arising from this submission.

\section*{Data Availability}
Hydrodynamical simulations used the {\sc phantom} code which is available from \url{https://github.com/danieljprice/phantom}. The input files for generating the SPH simulations and radiative transfer models will be shared on reasonable request to the corresponding author. Radiative transfer calculations were performed using {\sc mcfost} which is available on a collaborative basis from CP.



\bibliographystyle{mnras}
\bibliography{summaryCB} % if your bibtex file is called example.bib
%%%%%%
\appendix
%%%
\section{Sinusoid Fitting}
\label{app:sinefit}
\begin{figure}
    \centering
    \includegraphics[width=0.9\columnwidth]{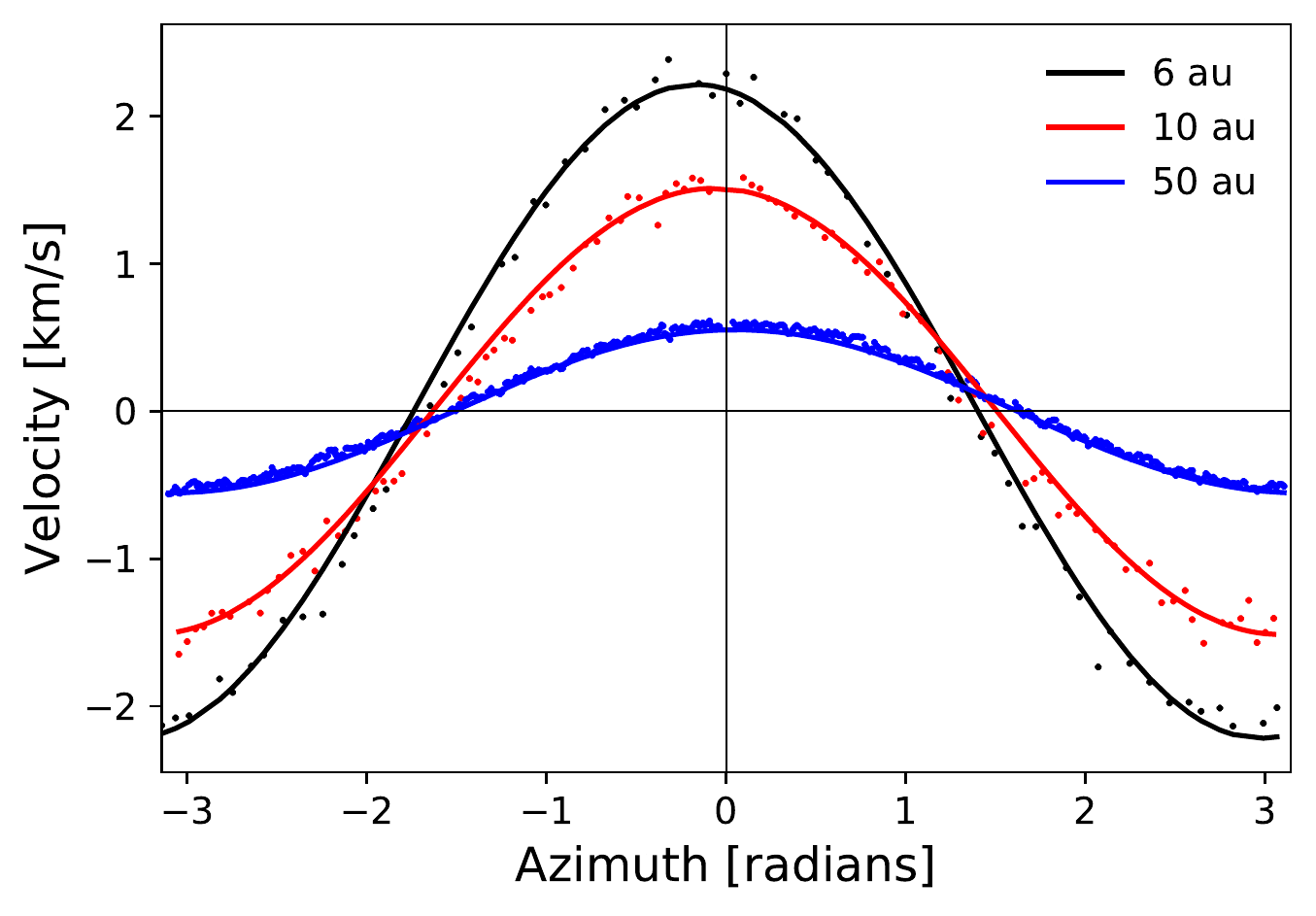}
    \caption{Velocity values extracted from annuli centered on three radii from the $v_{\rm{peak}}$ map of the $10^\circ$ binary inclination model viewed at 5$^\circ$. The curves are the resulting fits to the data points.The shift in position of the velocity peaks indicates the warp struture.}
    \label{fig:sinfit}
\end{figure}

We fit sinusoids to annuli of the disc to analyse the velocity maps as described in section \ref{sec:sinefit}. Here, we briefly justify this choice and show examples. Fitting annuli required that the disc is viewed face on so that a given annulus traces regions at equal radii. For the velocity maps produced for a nonzero viewing inclination, we first deproject the maps. Fig.~\ref{fig:sinfit} shows an example of the pixel data extracted from a velocity map in three annuli and the sinusoids fitted.

We see that the warp is naturally revealed in the phase shift of the sinusoids. The amplitude of the sinusoid provides a reliable measure of the peak velocity because the contribution from all azimuths has the effect of averaging out the noise. The fit is only meaningful in the warped region. For example, in some of the synthetic velocity maps there is a wave propagating outwards beyond the warped region and an annulus would contain mainly either red- or blue-shifted emission. Some care must be taken then to check that a sinusoid is a good fit to the data.

\section{Keplerian fitting with {\sc eddy}}
\label{sec:eddyfitting}
\begin{figure}
    \centering
    \includegraphics[width=\columnwidth]{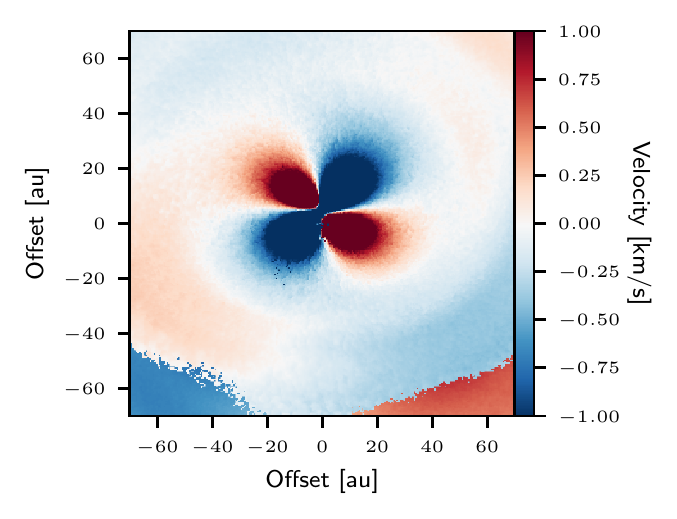}
    \caption{The residuals for the best fitting Keplerian model found with {\sc eddy} for the $10^\circ$ binary inclination model with a viewing inclination of 35${^\circ}$.}
    \label{fig:eddyfit}
\end{figure}

Fitting a Keplerian disc model to a warped disc is naturally going to produce a poor fit. However, a warped region may only affect the inner few au of an otherwise near-Keplerian disc. It is also a ubiquitous method for analysing spectral line observations of circumstellar discs. We therefore attempted to fit some example velocity maps from our modelling with {\sc eddy}.

The best fits were found by masking the inner region where the distortion was greatest and also by fixing the central mass and viewing inclination. An example of the residuals between the $v_{\rm{peak}}$ map and best fit model is shown in  Fig.~\ref{fig:eddyfit}. For this example, the fitting was applied to the annulus with inner and outer radii 37~au and 67~au. The mass, distance, $v_{\mathrm{lsr}}$ and inclination were fixed to the simulation parameters: 2~M$_\odot$, 140~pc, 0~km~s$^{-1}$ and 35$^\circ$ respectively. The central position and position angle were left as free parameters.

The residuals showed remarkably similar features for all inclinations 5-35$^\circ$. These comprise an 'X' shape in the centre and one or two spiral-like features just beyond. \citet{paneque2021} report this 'X' shape in the residuals form fitting Elias 2-27 with {\sc eddy}. This feature is the result of subtracting a Keplerian field from a warped velocity field. We suggest that these observations of Elias 2-27 are consistent with a central warp.

%TC:ignore
%\detailtexcount{summaryCB}
%TC:endignore
%%%%%%

% Don't change these lines
\bsp	% typesetting comment
\label{lastpage}
\end{document}